# Do peers share the same criteria for assessing grant applications?

Sven E. Hug[1,*] & Michael Ochsner[2]

[1] Department of Psychology, Social- and Business Psychology, University of Zurich, Binzmühlestrasse 14/13, 8050 Zurich, Switzerland
[2] FORS Swiss Centre of Expertise in the Social Sciences, University of Lausanne, Bâtiment Géopolis, 1015 Lausanne, Switzerland
* Corresponding author. Email: sven.hug@uzh.ch

**Abstract**

This study examines a basic assumption of peer review, namely, the idea that there is a consensus on evaluation criteria among peers, which is a necessary condition for the reliability of peer judgements. Empirical evidence indicating that there is no consensus or more than one consensus would offer an explanation for the *disagreement effect*, the low inter-rater reliability consistently observed in peer review. To investigate this basic assumption, we have surveyed all humanities scholars in Switzerland on 23 grant review criteria. We have employed latent class tree modelling to identify subgroups in which scholars rated criteria similarly (i.e. latent classes) and to explore covariates predicting class membership. We have identified two consensus classes, two consensus-close classes, and a consensus-far class. The consensus classes contain a *core consensus* (ten criteria related to knowledge gaps; feasibility; rigour; comprehensibility and argumentation; academic relevance; competence and experience of the applicant) and a *broad consensus* that includes the core consensus plus eight contribution-related criteria, such as originality. These results provide a possible explanation for the disagreement effect. Moreover, the results are consistent with the notion of conservatism, which holds that original research is undervalued in peer review, while other aspects, such as methodology and feasibility, are overweighted. The covariate analysis indicated that age and having tenure increases from the consensus-far to the consensus-close to the consensus classes. This suggests that the more academic experience scholars accumulate, the more their understanding of review criteria conforms to the social norm.

**Keywords**

peer review, grant funding, evaluation criteria, inter-rater reliability, disagreement effect, conservatism

# 1. Introduction

Scholars who review the research of their colleagues often arrive at very different judgements (Cicchetti, 1991; Lee et al., 2013). This phenomenon is so robust that it has been observed since the emergence of modern peer review in the late 1960s and early 1970s,[1] as evidenced, for example, by the studies of Scott (1974) and Kerzendorf et al. (2020). Low inter-rater reliability could therefore be considered a hallmark of peer review.[2] Despite the robustness and repeated replication[3] of this 'disagreement effect', few studies have investigated the reasons behind it (Bornmann, Mutz and Daniel, 2010; Seeber et al., 2021). It seems therefore that research on peer review focuses more on discovering and confirming effects and less on explaining effects and building theories. There are indeed some scholars who concluded that explanatory mechanisms need to be incorporated more in analyses of peer review (Johnson and Hermanowicz, 2017), that there are only folk theories on peer review that explain little (Reinhart, 2017), and that there is a general lack of theory in research on peer review (Hirschauer, 2004). In addition, Elson, Huff and Utz (2020) called for field experiments to increase knowledge on mechanisms and determinants of peer review processes. The aim of the present study is therefore to make an explanatory contribution to research on peer review. To this end, we empirically examine whether scholars share the same evaluation criteria, a basic assumption of peer review, and propose what this basic assumption could enable and explain with respect to peer review.

## 1.1 Criteria consensus – A basic assumption of peer review

According to Henkel (1998), one of the basic assumptions of all academic peer review procedures is that there is a general consensus on and understanding of what the key evaluation criteria within a discipline are. Other authors make similar assumptions with regard to journal peer review (Arvan, Bright and Heesen, 2020; Hirschauer, 2005, 2010) and grant peer review (Feller, 2013; Forscher et al., 2019). The main assumption put forward by these authors could be summarized as follows: peer review is based, among other things, on the idea that there is a consensus on evaluation criteria among peers. Since the authors each conceptualize the assumption in a slightly different way, the following points require clarification. First, the authors use the term peer differently. While some understand peers as the members of a discipline or field (e.g. Henkel, 1998), others construe peers as readers and reviewers of a journal (e.g. Arvan et al., 2020). To take such differences into account, we conceive peers as an 'elastic term' (Henkel, 1998, p. 293) that can refer to any group of researchers, ranging from the most knowledgeable scholars doing exactly the same kind of research to the members of a discipline to a broad assemblage of scholars from different fields to the scholarly community in general. To indicate the elastic and gradual nature of the notion of peer, we propose to speak of the *degree of peerness*. Second, some authors assume that there is a consensus on criteria among peers regardless of a particular review situation (e.g. Hirschauer, 2005, 2010), whereas others focus on the particular application of criteria and suppose that peers use the same criteria when assessing the very same manuscript or grant proposal (e.g. Forscher et al., 2019). Hence, the assumption can either refer to criteria that peers prefer in general (e.g. a criteria consensus in a scholarly community), or to those they apply in an actual review situation. The latter version of the assumption is contained in the ideal of impartiality, which implicitly underlies quantitative research on bias in peer review (Lee et al., 2013). The ideal of impartiality requires that, among other things, reviewers have to 'interpret and apply evaluative criteria in the same way in the assessment of a submission' (Lee et al., 2013, pp. 3–4). Third, there are differences in terms of what exactly has to be shared among peers. While some authors assume that criteria must be shared (Forscher et al., 2019; Henkel, 1998), others, more broadly, assume a common understanding of what constitutes quality, merit, excellence or good science (Arvan et al., 2020; Feller, 2013; Hirschauer, 2005, 2010). Lastly, the authors do not assume that perfect agreement, complete consensus, or unanimity is necessary with regard to the evaluation criteria. Rather, partial agreement or some degree of consensus seems sufficient. For example, Arvan et al.

(2020, p. 4) presuppose that 'there exists broad (if rough) intersubjective agreement', Henkel (1998) assumes a consensus on the key evaluation criteria, and Forscher et al. (2019) speak of agreeing on similar criteria. In the context of the assumption, we understand partial agreement and consensus as 'a high proportion of peers approve a set of evaluation criteria'. Note that we will examine one particular version of the basic assumption in this study: we will define peers as the community of humanities scholars in Switzerland, assume that evaluation criteria have to be shared, and focus on grant review criteria that scholars prefer in general and not those they apply in a particular review situation.

1.2 Reasons for the basic assumption

While it is clear from the literature that some authors consider a consensus on criteria to be a basic assumption of peer review, the reasons for this assumption are far less evident, as only Arvan et al. (2020) and Forscher et al. (2019) explain why a consensus is essential and what it is supposed to enable. Arvan et al. (2020) argue that only if reviewers and readers of a journal agree on what constitutes quality, the judgements of reviewers are useful to readers when deciding how to allocate their time among scientific journals. Forscher et al. (2019) contend that if reviewers do not agree on their criteria, this will result in arbitrary and unreliable judgements. In the following, we propose further reasons why the assumption of a consensus on criteria could be relevant to peer review and what a consensus could enable.

According to Thorngate, Dawes and Foddy (2009), applicants, observers, and judges consider judgements of merit to be fair if the judgements are valid, consistent, efficient, equitable, and transparent. We argue that a consensus on criteria among peers contributes, to varying degrees, to the validity, consistency, and efficiency of peer review. Most importantly, a consensus contributes to validity. Thorngate et al. (2009) state that fair judgements use criteria deemed valid by a high proportion of applicants, observers, and judges. This corresponds to what we termed 'partial agreement' and 'consensus' in the discussion of the basic assumption above.

Furthermore, shared criteria enable and facilitate the consistency of reviews across reviewers and over time as they provide a common and stable frame of reference. This argument on consistency is in line with the argument of Forscher et al. (2019) on arbitrariness and reliability (see above). Lastly, a consensus on criteria makes peer review more efficient since the applicants as well as the reviewers do not need to define and negotiate the criteria completely anew each time and are therefore faster in crafting applications and writing reviews. Shared criteria thus contribute to the validity, consistency, and efficiency of peer review, which in turn, according to Thorngate et al. (2009), promote the perceived fairness of merit judgements. Since what Thorngate et al. (2009) describe as fairness is commonly referred to as legitimacy,[4] we conclude that a consensus on criteria promotes the legitimacy of peer review.

Mutually shared knowledge facilitates communication (Holler and Wilkin, 2009) and socially shared preferences, motives, norms, cognitions, etc. make group information processing possible (Tindale and Kameda, 2000). Recent studies have illustrated the importance of establishing a common understanding of criteria as a basis for interaction in panel peer review (Derrick, 2018; Derrick and Samuel, 2017; Pier et al., 2017). We therefore propose that a consensus on criteria provides a common conceptual and semantic ground that enables and facilitates the interaction, communication, and information processing of peers involved in an assessment (authors, applicants, reviewers, panellists, editors, etc.).

Based on the politeness model of Brown and Levinson (1987), Hyland and Diani (2009) argue that reviews are fraught with potential face threats for both reviewers and those whose work is being reviewed. Hyland and Diani (2009) maintain that, despite the many threats, reviews succeed because those involved are oriented towards disciplinary evaluative standards and norms of engagement. While Hyland and Diani (2009) focus on linguistic strategies, such as hedging and stancetaking, we propose that a consensus on criteria is another important factor in mitigating face-threatening acts. If applicants and reviewers adhere to shared criteria, face-threatening acts are less likely to occur and less severe. For example, when a reviewer assesses an application negatively and rejects it based on shared criteria, negative reactions from the applicant and observing peers will be less likely because the reviewer used valid criteria and thereby demonstrated that she/he is a qualified member of a scholarly community.

Who is a peer and what is a peer are perennial questions in peer review practice and research. In the literature, subject matter expertise is considered a key attribute of reviewers (e.g. Chubin, 1994; Gallo, Sullivan and Glisson, 2016; Steiner Davis et al., 2020). Less widespread, for example, is the view that peers are defined by reputation, standing, and eliteness (e.g. Ravetz, 1971; The British Academy, 2007). We propose that review criteria provide another dimension for determining the degree of peerness (i.e. who is considered a peer). Specifically, if there exists a set of shared criteria within a scholarly community, the degree to which individuals approve and apply these criteria is indicative of their membership in that community. Take, for example, a professor who has considerable subject matter expertise and an excellent academic reputation, who has recently retired, and now works for an ideological think tank from time to time. This professor emeritus would only be considered a peer if she/he still approved and applied the criteria of her/his discipline – and not those of the think tank or the political sphere. In this way, a consensus on criteria and the degree to which a reviewer is committed to it indicates how legitimate she/he is to make a judgement as a peer and to what extent this judgement represents the judgement of a scholarly community.

So far, we have assumed that a *single* consensus on criteria is desirable and beneficial to peer review, and we have thus discussed what a consensus might enable. However, a single consensus may not always be desirable. In cases where diversity of opinion among reviewers is favoured (e.g. Harnad, 1979; Lee, 2013), it seems more desirable that reviewers prefer, apply, and emphasize different criteria to facilitate complementary assessments. For instance, societal relevance may not be part of a criteria consensus among basic researchers, while it may be included in a consensus among applied researchers. Hence, a funding agency could assemble a panel representing basic and applied researchers to ensure that the potential for societal benefit of basic research projects is also assessed.

1.3 Research questions

We have noted above that some authors consider a criteria consensus to be a basic assumption of peer review, and we have proposed several reasons why a consensus could be beneficial to peer review and what a consensus could enable. Since this assumption has not been empirically investigated (Arvan et al., 2020), we explored whether there is a consensus on criteria for assessing grant proposals of early-career researchers among humanities scholars in Switzerland.[5] Specifically, we conducted a survey on 23 evaluation criteria, applied latent class tree modelling (van den Bergh and Vermunt, 2019) to identify subgroups in which scholars rated criteria similarly (i.e. latent classes), and explored covariates to predict class membership (age, gender, research field, tenure, international orientation, experience as grant reviewer, application of a doctoral student or a postdoctoral researcher). Based on the latent classes, we then examined whether there is a consensus on criteria among all scholars or whether there are different consensuses in subgroups of scholars. If the same criteria – and only these criteria – are rated positively in each latent class and to the same degree, this suggests that there is a consensus among all scholars. If, however, there are latent classes in which different criteria are rated positively, this suggests that there are several consensuses. Accordingly, this study addresses the following research questions. (1) Are there subgroups in which scholars rate evaluation criteria similarly (i.e. latent classes)? (2) Is there one consensus on evaluation criteria or are there several consensuses? (3) Are some of the covariates predictive of scholars' class membership?

As we aim to provide an explanatory contribution to research on peer review with this study but have not yet stated what we seek to explain, we shall do so now. As noted above, a consensus on criteria enables a high inter-rater reliability in peer review. Conversely, if there is no consensus – or more than one consensus – this will result in low inter-rater reliability because reviewers use different criteria. Hence, empirical evidence indicating that there is no consensus or more than one consensus on criteria would offer an explanation for the *disagreement effect*, the low inter-rater reliability observed in peer review. Naturally, high or low inter-rater reliability is not caused by shared or unshared criteria alone (for other factors, see Brezis and Birukou, 2020; Lee, 2013; Lee, 2015; Sattler et al., 2015; Seeber et al., 2021). Furthermore, a consensus in a scholarly community does not guarantee that the shared criteria are actually applied in a review situation. For example, the criteria prescribed by a journal or funding agency could constrain scholars from using criteria shared in their community. However, reviewers' adherence to reviewer guidelines seems to be rather weak. Eve et al. (2021) found that 77% of the analysed review reports from PLOS ONE contained comments on the novelty and significance of a paper, although the journal asks reviewer specifically not to use these criteria. In addition, the journal requires reviewers to remark upon reproducibility, but Eve et al. (2021) found they rarely do so. Langfeldt (2001, p. 835) concluded with regard to peer review at the Research Council of Norway that 'the guidelines given to the panels had little effect on the criteria they emphasized, whereas mail reviewers were more consciously attempting to write reviews in accordance with the guidelines.' Reale and Zinilli (2017) demonstrated that project funding criteria imposed by the Italian Ministry for Education and Research had little to no impact on peer review practices. Moreover, research on group decision-making has shown that task-relevant preferences and cognitions 'shared among most or all of the group members exert an extraordinary influence on group decision processes and outcomes' (Tindale, Kameda and Hinsz, 2003, p. 381).[6] For these reasons, we can expect that a criteria consensus in a scholarly community influences a review situation.

## 2. Methods

2.1 Data

Based on studies that explored and mapped humanities scholars' understanding of research quality (Hug, Ochsner and Daniel, 2013; Ochsner, Hug and Daniel, 2013), we designed a questionnaire comprising 23 criteria for assessing grant applications from the humanities. Some of the criteria were grouped into broader themes (originality, rigour, relevance, feasibility, applicant). Participants were asked to rate the criteria on a six-point scale labelled as follows: (1) strongly disagree, (2) disagree, (3) slightly disagree, (4) slightly agree, (5) agree, (6) strongly agree. We created two versions of the questionnaire, which contained the same evaluation criteria, but had different introductions. Participants were either told that they were rating criteria of a career funding scheme for doctoral students or for postdoctoral researchers, such as Doc.CH or Ambizione of the Swiss National Science Foundation (SNSF). The survey also included four questions on personal characteristics (age, tenure, orientation of research, experience as grant reviewer). It was drafted in German and translated to English, French, and Italian based on the TRAPD method (Harkness, 2003). The criteria included in the survey are much more detailed than criteria prescribed by the SNSF. On a general level, the criteria in the survey and those of the SNSF overlap to some extent, which is typical of criteria used or preferred by researchers and criteria prescribed by funding agencies (Hug and Aeschbach, 2020). The survey questions as well as the SNSF criteria are listed in the Supplementary Material (Tables 1 and 3).

The Swiss Higher Education Information System (SHIS) provides a taxonomy of academic fields and disciplines that we used to identify humanities scholars. Specifically, we included scholars from three of the four humanities fields (languages and literatures, history and cultural sciences, law), and we excluded scholars from the field of religious studies and theology because other researchers conducted a national survey in this field at the same time. We searched the websites of all Swiss universities and collected email address, discipline, and gender of a scholar if three conditions were met: affiliated with a Swiss university, holding at least a doctoral degree, belonging to a humanities discipline/field (SHIS taxonomy). In case of missing data or uncertainty we contacted the respective university or queried a database of the Rectors' Conference of the Swiss Universities containing all scholars with a habilitation employed at a Swiss university per SHIS field. Based on their discipline, we assigned the scholars to one of the three fields, and we randomly assigned the scholars to the questionnaire either on criteria for applications of doctoral students or on criteria for postdoctoral researchers. The survey was administered online to 2,609 scholars in 2015 (November to December), 938 of whom completed it, which corresponds to a response rate of 36%. While the response rates to the questionnaires on the criteria for applications of doctoral students and postdoctoral researchers were the same (37% and 35%, respectively), they differed somewhat with regard to gender (women 40%, men 34%). Women are thus slightly overrepresented among the respondents. The response rate in the field of law (25%) was considerably lower than in the fields of languages and literatures (39%) and history and cultural sciences (42%). We suppose that this low response rate is due to the significant number of practitioners who teach law at Swiss universities. In fact, we received many emails from lawyers who stated that they considered it inappropriate to answer the questionnaire because they were not much involved in grant funding and academic life anymore. Overall, the percentage of missing values was very low (1.2%) and particularly low among the 23 evaluation criteria (0.8%). More data was missing in the answers to the four questions on personal characteristics (3.5%).

## 2.2 Statistical analysis

Latent class analysis (LCA) is a popular method for clustering individuals into homogeneous subgroups (Masyn, 2013; Petersen, Qualter and Humphrey, 2019). However, when using LCA for exploratory purposes, it is often difficult to decide on the number of classes to retain in the analysis. In addition, LCA of a large data set often results in a large number of classes that are difficult to interpret substantively. To address these issues, van den Bergh, Schmittmann and Vermunt (2017) developed the latent class tree (LCT) procedure that imposes a hierarchical structure on latent classes. The procedure starts with an initial set of nodes (i.e. classes), and then estimates 'a 1- and a 2-class model for the subsample at each node of the tree. If a 2-class model is preferred according to the fit measure used, the subsample at the node concerned is split and two new nodes are created. The procedure is repeated at the next level of the hierarchical structure until no further splits need to be performed' (van den Bergh et al., 2017, pp. 14-15). The LCT method uses the Bayesian information criterion (BIC) as fit measure and splits classes 'as long as the difference between the BIC of the estimated 1- and 2-class models, $\Delta BIC = BIC(1) - BIC(2)$, is larger than 0' (van den Bergh et al., 2017, p. 16). Due to the hierarchical nature of the LCT approach, classes can be interpreted at any level of the tree, and the number of classes can thus be determined based on substantive interpretation (van den Bergh, van Kollenburg and Vermunt, 2018). While classes at higher levels of the tree represent more dominant differences between individuals, lower-level classes reveal more fine-grained differences (van den Bergh and Vermunt, 2019).

We analysed the survey data in R 3.5.2 (R Core Team, 2018) using the LCTree package of van den Bergh (https://github.com/MattisvdBergh/LCT) and in Latent Gold 5.1 (Vermunt and Magidson, 2016). The LCTree package provides functions to perform a bias-adjusted three-step analysis (Vermunt, 2010) of LCTs, that is, it estimates hierarchical latent classes, assigns respondents to classes and quantifies classifications errors, and predicts class membership based on covariates while correcting for classification errors. The LCTree package accesses Latent Gold to estimate and predict LCT models. Our analysis included the following main steps. First, we estimated a standard LCA with 1 to 10 classes in Latent Gold using the default settings and handling missing data full information maximum likelihood. Based on these results, we calculated the *relative improvement of fit measure* proposed by van den Bergh et al. (2018) for the BIC to determine the number of classes at the root node of the LCT (i.e. the initial number of classes from where the LCT starts). Second, we estimated an LCT with the LCT function of the LCTree package and used the BIC as split criterion. Third, we predicted class membership based on the covariates with the exploreTree function of the LCTree package and used the maximum likelihood method for bias adjustment as proposed by Vermunt (2010) and Bakk, Tekle and Vermunt (2013). The R code used in the second and third step of this analysis is provided in the Supplementary Material. Fourth, we conducted a covariate analysis in Latent Gold for each split of the LCT as the results obtained with the exploreTree function suggested possible interactions between some covariates and as interactions of covariates cannot be defined in the exploreTree function. Specifically, we ran the Step3 module in Latent Gold with the same parameters as the exploreTree function did and added interactions between the covariates age and tenure as well as age and experience as grant reviewer. As Latent Gold cannot handle interactions between continuous and nominal variables, we calculated quintiles for the covariate age and split the sample into five age groups, each comprising approximately 20% of the respondents. Note that the LCTree package handles missing values of response variables (i.e. the 23 evaluation criteria) by full information maximum likelihood and excludes cases listwise if there are missing values in the covariates. We thus included all respondents in the analysis who had complete data in the covariates and allowed missing values in the response variables (n = 866). We also conducted a sensitivity analysis using complete cases only (i.e. no missing values in covariates and response variables) and report the results of this analysis in the Supplementary Material (n = 760). The results of the two analyses are consistent.

# 3. Results

## 3.1 Latent class tree

The relative improvement of the BIC ($RI_{BIC}$) indicates that the improvement is marginal for models with more than four classes (see Supplementary Material, Table 5), and therefore the number of classes at the root node of the LCT was set to four. Figure 1 shows the result of the LCT analysis, a tree with four main latent classes (LC 1, 2, 3, 4) and four subclasses (LC 1.1, 1.2, 3.1, 3.2). While the size of five classes can be considered sufficiently large, three classes are small, each containing around 10% of the respondents (LC 3.1, 3.2, 4). LC 3.2 is particularly small (7.6%), suggesting that LC 3 should not be split further. The inspection of ΔBIC of each split shows that the expansion of the tree from one to four classes yields a very large improvement (ΔBIC = 2'824), while splitting LC 1 (ΔBIC = 80) and LC 3 (ΔBIC = 43) yields marginal improvements. As splitting LC 3 produces two small subclasses, yields only a small improvement of the BIC, and generates subclasses that are not of substantive interest (see below), we decided not to split LC 3 and to retain five classes in the final tree (LC 1.1, 1.2, 2, 3, 4).

Figure 1. Structure of the latent class tree (n = 866). Class size is indicated in percent of the sample size.

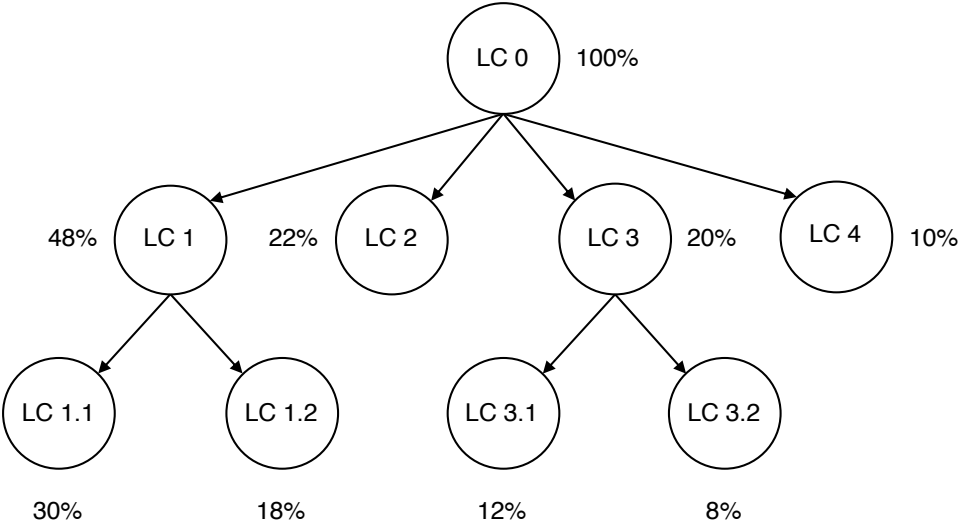

The main classes generally differ from each other with respect to the average ratings of the 23 evaluation criteria (see Figure 2 and Wald statistics in the Supplementary Material, Table 6). There are only three criteria where visual inspection and Wald tests indicate that the ratings of several classes overlap: 'independence' (LC 1, 2, 4), 'variety' (LC 1, 2, 4), and 'letter of recommendation' (LC 1, 2; LC 2, 3). In addition, there is some overlap between LC 1 and 2 (8 of 23 criteria) as well as between LC 2 and 3 (11 of 23 criteria). Note that although LC 2 and 3 overlap the most, they differ considerably with regard to most of the originality criteria ('new approach', 'new topic', 'innovative data', 'new paradigm', 'new findings') as well as with regard to 'societal relevance' and 'cultural heritage'. The Wald tests also indicate that LC 2 and 3 differ with regard to further criteria ('independence', 'gaps', 'academic relevance', 'complexity', 'variety'), but these differences are less pronounced (Figure 2). In accordance with the LCT method, the overlap between subclasses is much larger than the overlap between main classes, that is, each pair of subclasses overlaps on 15 of 23 criteria (see Figure 3 and Wald statistics in the Supplementary Material, Table 7 and 8). Although visual inspection and Wald tests indicate differences between LC 3.1 and 3.2 regarding the rating of eight criteria, we consider these differences not to be meaningful with regard to the research questions because the respective criteria receive maximal ratings in LC 3.2 and very high ratings in LC 3.1, which means the same: both classes agree on the same criteria. We thus decided not to split LC 3 into LC 3.1 and 3.2. Accordingly, the final tree includes two classes with very high average ratings (LC 2, 3), two classes with high ratings (LC 1.1, 1.2), and a class with moderate ratings (LC 4). With respect to research question 1, this suggests that there are subgroups in which scholars rate evaluation criteria similarly and that the ratings generally differ between subgroups.

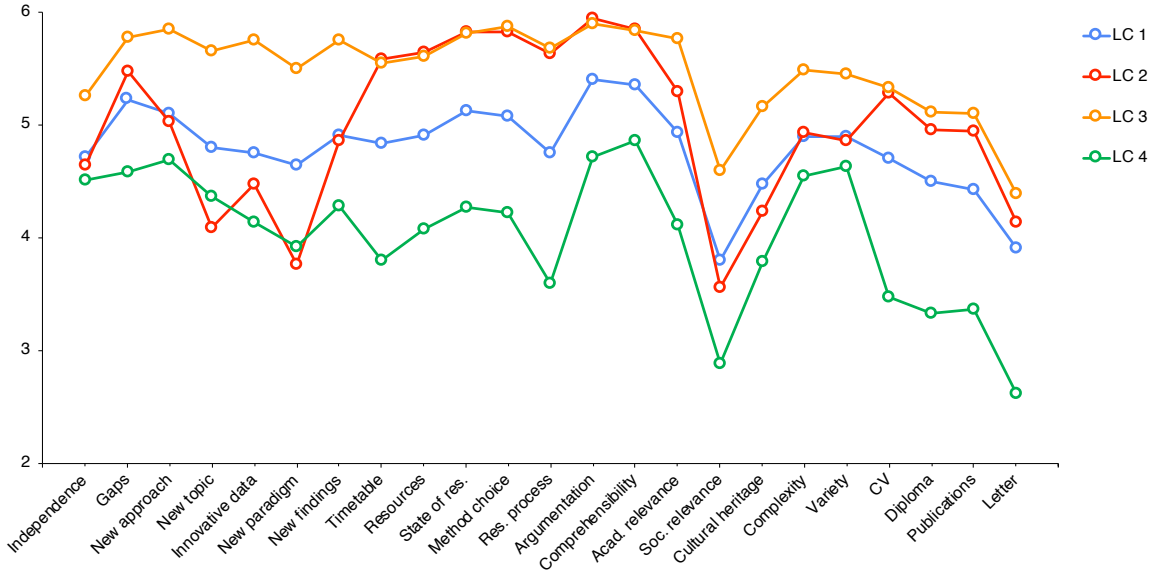

Figure 2. Average rating of the criteria by main classes of the latent class tree (n = 866) on a six-point scale (1 strongly disagree, 2 disagree, 3 slightly disagree, 4 slightly agree, 5 agree, 6 strongly agree).

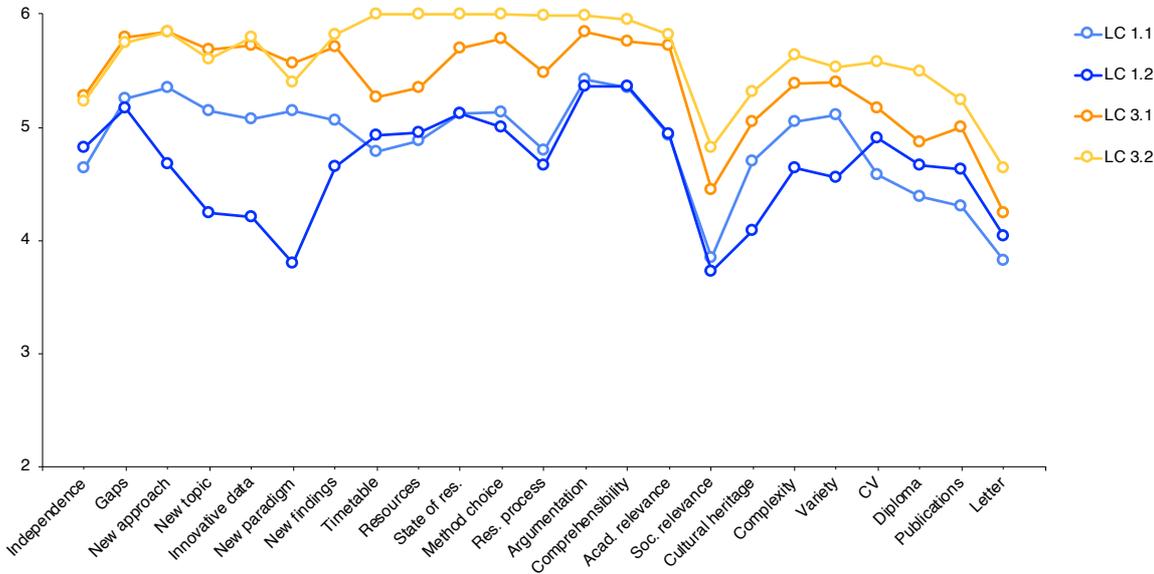

Figure 3. Average rating of the criteria by subclasses of the latent class tree (n = 866) on a six-point scale (1 strongly disagree, 2 disagree, 3 slightly disagree, 4 slightly agree, 5 agree, 6 strongly agree).

Having identified different subgroups, the question arises whether these subgroups represent different consensuses (research question 2). Note that we defined a consensus as 'a high proportion of peers approving a set of evaluation criteria'. To visualise consensuses in the final tree that fit this definition, we used the probabilities of rating a criterion as '5 – agree' or '6 – strongly agree'. According to Figure 4, scholars in LC 2 and 3 have a very high probability to agree or strongly agree with many of the evaluation criteria. In fact, the probability in these classes is close to 1 for some criteria. As noted above, in LC 2 and 3, some criteria are rated identically, while other criteria are rated very differently (e.g. criteria related to originality). We thus consider LC 2 and 3 each to represent a distinct *consensus class*. Furthermore, the rating patterns of LC 2 and 1.2 are very similar, but the probability to agree or to strongly agree is generally lower in LC 1.2 (high probability) than in LC 2 (very high probability). The same applies to LC 3 and 1.1. The two consensus classes are thus mirrored at a lower probability level in LC 1.1 and 1.2. To indicate this, we refer to LC 1.1 and 1.2 as being *consensus-close*. Lastly, LC 4 features some of the patterns of the other four classes. For example, while 'societal relevance' and 'letter' are rated lowest, 'argumentation' and 'comprehensibility' are rated highest, and the ratings of applicant-related criteria decrease from 'CV' to 'letter'. However, the probability to agree or strongly agree with the 23 criteria is moderate to low in LC 4 and therefore considerably lower than in the other classes. We thus refer to LC 4 as being *consensus-far*.

To characterize the substantive content of the two consensuses (i.e. sets of criteria approved by a high proportion of scholars in LC 2 and 3), we set a probability threshold in Figure 4. Naturally, this threshold of 0.85 is, like any such threshold, arbitrary, and, depending on the threshold, more or less criteria are included in the core and broad consensus. Drawing on the criteria that are above the threshold of 0.85 the substantive content of the two consensus classes, can be characterized as follows. Both LC 2 and 3 share a core set of ten criteria that assess if knowledge gaps are identified in the application ('gaps'), if the project is feasible ('timetable', 'resources'), if it is rigorous ('state of research', 'method choice', 'research process') and well presented ('argumentation, 'comprehensibility'), if it is relevant to the scholarly community ('academic relevance'), and if the applicant is competent and experienced ('CV'). While the criteria of LC 2 above the threshold are identical to this core set, LC 3 includes eight additional criteria above the threshold. These criteria assess the specific contributions and outcomes of a research project in terms of originality ('new approach', 'new topic', 'innovative data', 'new paradigm', 'new findings') as well as the project's wider contribution to a research field ('complexity', 'variety'), and if the project is independently designed and conducted by the applicant ('independence'). In summary, we suggest a *core consensus* consisting of ten criteria (LC 2) and a *broad consensus* that includes the core consensus plus eight additional criteria (LC 3). Note that although the threshold we set is arbitrary, the main findings do not change when the threshold is changed or removed: there is a set of criteria that is rated equally positively in both consensus classes ('timetable', 'resources' etc.); criteria related to originality are rated very positively in one consensus class (LC 3), while they are rated considerably lower in the other consensus class (LC 2).

Figure 4. Probability of rating a criterion as 5 (agree) or 6 (strongly agree) by classes of the final latent class tree (n = 866). The dashed line indicates a probability of 0.85.

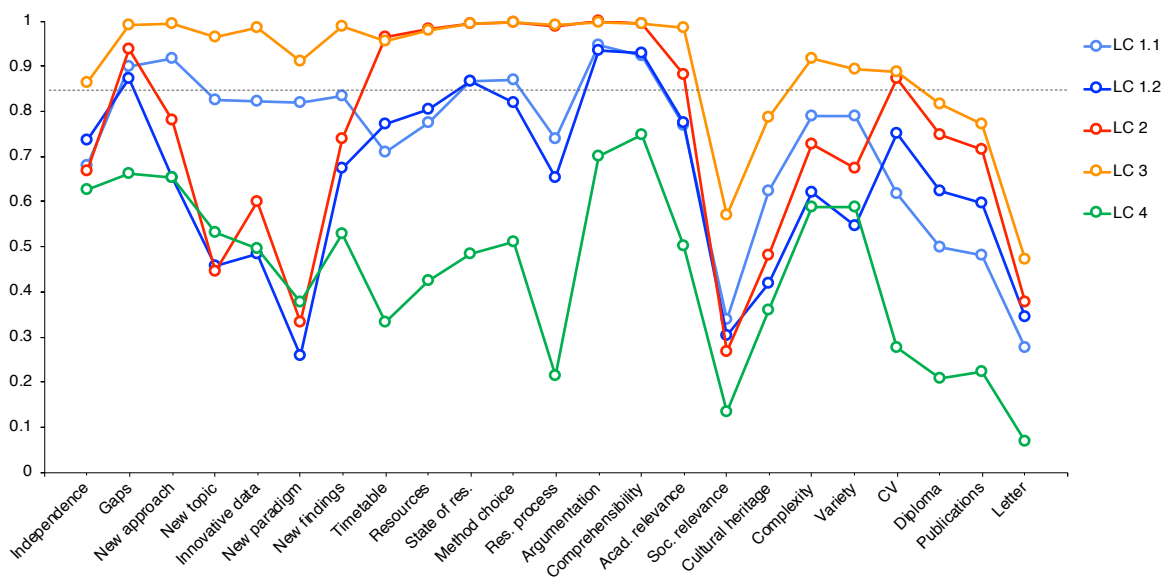

## 3.2 Covariate analysis

The class specific probabilities of the covariates (Table 1) suggest that gender, age, and tenure are predictive of membership in the main classes. Specifically, age and having tenure increases among the main classes (LC 4 < LC 1 < LC3 < LC 2), suggesting that academic experience influences whether scholars are consensus-far (LC 4) or consensus-close (LC 1), share a broad consensus (LC 3) or a core consensus (LC 2). The effect of academic experience is partly supported by the probability of having experience as grant reviewer that decreases from LC 2 (0.71) to LC 3 (0.61) to LC 1 and 4. Note that there is no decrease between LC 1 (0.48) and LC 4 (0.47). In contrast to age and tenure, the class specific probabilities of gender show a less clear pattern. The percentage of women is highest in LC 3 (0.48), lowest in LC 4 (0.30), and identical in LC 1 and LC 2 (0.41). This might indicate that women prefer a broader set of criteria to evaluate grant proposals (LC 3) or that (young and untenured) men are less conscious of criteria norms (LC 4). Furthermore, Table 1 suggests that the covariate 'research field' is predictive of membership in the subclasses LC 1.1 and 1.2. The percentage of scholars from the field of languages and literatures is considerably higher in LC 1.1 (0.50) than in LC 1.2 (0.31), while the percentage of law scholars is twice as high in LC 1.2 (0.27) than in LC 1.1 (0.14). This field effect could offer an explanation for the difference between the core consensus (LC 2) and the broad consensus (LC 3), as the rating patterns of LC 2 and 1.2 as wells as LC 3 and 1.1 are very similar. However, the research field is not related to the main classes as the respective class specific probabilities are virtually identical. We speculated above that women might prefer a broader set of criteria (LC 3), and indeed we find this tendency also among the subclasses as the percentage of women in LC 1.1 (0.45), the subclass with more positively rated criteria, is higher than in LC 1.2 (0.35). Lastly, the class specific probabilities do not indicate that the covariates 'orientation of research' (national, international, both) and 'application of doctoral student or postdoc' predict membership in the main classes or subclasses (Table 1).

Table 1. Relative frequency of the covariates in the sample (n = 866) and class specific probabilities of the covariates.

| Covariates | Sample | Main classes | | | | Subclasses | |
| --- | --- | --- | --- | --- | --- | --- | --- |
| | | LC 1 | LC 2 | LC 3 | LC 4 | LC 1.1 | LC 1.2 |
| Gender | | | | | | | |
|    Men | .59 | .59 | .59 | .52 | .70 | .55 | .65 |
|    Women | .41 | .41 | .41 | .48 | .30 | .45 | .35 |
| Criteria for applications of | | | | | | | |
|    Doctoral students | .53 | .53 | .52 | .50 | .59 | .50 | .58 |
|    Postdoctoral researchers | .47 | .47 | .48 | .50 | .41 | .50 | .42 |
| Field | | | | | | | |
|    Languages and literatures | .41 | .43 | .40 | .37 | .41 | .50 | .31 |
|    History and cultural sciences | .41 | .39 | .43 | .45 | .39 | .37 | .43 |
|    Law | .18 | .19 | .17 | .18 | .20 | .14 | .27 |
| Age | | | | | | | |
|    28-37 | .19 | .22 | .15 | .13 | .22 | .19 | .25 |
|    38-43 | .21 | .22 | .09 | .21 | .40 | .20 | .24 |
|    44-49 | .21 | .19 | .29 | .23 | .11 | .21 | .16 |
|    50-57 | .20 | .17 | .23 | .27 | .18 | .17 | .18 |
|    58-82 | .19 | .20 | .25 | .17 | .09 | .22 | .17 |
| Tenure | | | | | | | |
|    Yes | .39 | .35 | .54 | .40 | .26 | .34 | .38 |
|    No | .61 | .65 | .46 | .60 | .74 | .66 | .62 |
| Orientation of research | | | | | | | |
|    Mainly national | .14 | .18 | .09 | .10 | .13 | .14 | .22 |
|    Mainly international | .41 | .38 | .44 | .43 | .40 | .38 | .38 |
|    Equally national and international | .45 | .44 | .47 | .47 | .47 | .48 | .40 |
| Experience as grant reviewer | | | | | | | |
|    Yes | .56 | .48 | .71 | .61 | .47 | .47 | .54 |
|    No | .44 | .52 | .29 | .39 | .53 | .53 | .46 |

As we do not have a random sample at hand, we based our interpretation of the covariate analysis on class specific probabilities (effect sizes) and substantial meaning rather than on Wald tests (inferential statistics). However, the results pointed out above are also those that would be highlighted when examining the Wald tests ($p < .05$) shown for documentation purposes in Tables 2 and 3: The main effects of gender, age, and tenure as well as the interaction of age and tenure are significantly linked to membership in the main classes (Table 2). In particular, the main classes LC 1, 2, and 3 differ significantly from LC 4 regarding gender, age, tenure, and the interaction of age and tenure (Table 3). Furthermore, the research field is significantly linked to membership in the subclasses LC 1.1 and 1.2, whereas gender is not (Table 2).

Table 2. Wald statistics of the covariates by main classes (LC 1, 2, 3, 4) and subclasses (LC 1.1, 1.2) of the final latent class tree (n = 866).

| Covariates | Main classes Wald | p | Subclasses Wald | p |
|---|---|---|---|---|
| Main effects | | | | |
| Gender | 10.59 | .02 | 1.21 | .27 |
| Application of doc or postdoc | 2.10 | .55 | 0.48 | .49 |
| Field | 3.57 | .74 | 6.92 | .03 |
| Age | 33.69 | <.001 | 4.13 | .39 |
| Tenure | 2492.90 | <.001 | 0.06 | .81 |
| Orientation of research | 7.89 | .25 | 0.51 | .78 |
| Experience as grant reviewer | 2.23 | .53 | 2.70 | .10 |
| Interaction effects | | | | |
| Age x Tenure | 3192.32 | <.001 | 3.33 | .50 |
| Age x Experience as grant reviewer | 16.43 | .17 | 2.30 | .68 |

Table 3. Wald statistics of the paired comparisons of the main classes LC 1, 2, 3, and 4 for the covariates gender, age, tenure and the interaction of age and tenure (n = 866).

| Paired comparison | Gender Wald | p | Age Wald | p | Tenure Wald | p | Age x Tenure Wald | p |
|---|---|---|---|---|---|---|---|---|
| LC 1 and 2 | 0.60 | .44 | 4.20 | .38 | 0.07 | .80 | 3.07 | .55 |
| LC 1 and 3 | 2.74 | .10 | 6.96 | .14 | 1.55 | .21 | 5.15 | .27 |
| LC 1 and 4 | 4.58 | .03 | 14.75 | .01 | 2491.58 | <.001 | 2868.56 | <.001 |
| LC 2 and 3 | 0.46 | .50 | 3.48 | .48 | 0.47 | .49 | 5.14 | .27 |
| LC 2 and 4 | 6.29 | .01 | 20.58 | <.001 | 695.36 | <.001 | 692.83 | <.001 |
| LC 3 and 4 | 9.80 | <.002 | 21.25 | <.001 | 944.30 | <.001 | 985.04 | <.001 |

## 4. Discussion and conclusion

Research on peer review mainly discovers and confirms effects and rarely attempts to explain them. In this study we therefore focused on a basic assumption of peer review, the idea that there is a consensus on evaluation criteria among peers, which is a necessary condition for the reliability of peer judgements. Empirical evidence indicating that there is no consensus or more than one consensus on criteria would offer an explanation for the *disagreement effect*, the low inter-rater reliability consistently observed in peer review.

We examined one version of the basic assumption: we defined peers as the community of humanities scholars in Switzerland, assumed that evaluation criteria have to be shared, and focused on grant review criteria that scholars prefer in general and not those they actually apply in a particular review situation. Specifically, we conducted a survey on 23 criteria among all humanities scholars affiliated with a Swiss university and employed latent class tree modelling (van den Bergh and Vermunt, 2019) to analyze the survey responses. The 23 criteria were derived from studies that mapped humanities scholars' understanding of research quality. We addressed three research questions: (1) Are there subgroups in which scholars rate evaluation criteria similarly (i.e. latent classes)? (2) Is there one consensus on evaluation criteria or are there several consensuses? (3) Are some characteristics of scholars (i.e. covariates) predictive of class membership?

We identified five latent classes: two consensus classes, two consensus-close classes, and a consensus-far class. The probability to agree with a criterion generally decreases from the consensus classes (very high) to the consensus-close classes (high) to the consensus-far class (moderate). We observed similar patterns in each of the five classes. For example, 'societal relevance' and 'letter of recommendation' were rated lowest, while 'argumentation' and 'comprehensibility' were rated highest. Furthermore, we observed that the rating patterns of the two consensus classes are very similar to those of the two consensus-close classes, but, as mentioned above, the probability to agree with a criterion is generally lower in the consensus-close classes. The two consensus classes contain a *core consensus* consisting of ten criteria and a *broad consensus* that includes the core consensus plus eight additional criteria. The core consensus comprises criteria that assess if knowledge gaps are identified in the proposal, if the project is feasible, if it is rigorous and well presented, if it is relevant to the scholarly community, and if the applicant is competent and experienced. The broad consensus additionally comprises criteria that assess the specific contributions and outcomes of a research project in terms of originality (e.g. new approach, new data, new findings) as well as the project's wider contribution to a research field (e.g. by showing the complexity of a topic, by conducting research outside the mainstream), and if the applicant makes an own, independent contribution. The additional criteria included in the broad consensus focus on *contribution-related* aspects of the application and applicant.

These results provide empirical evidence against the basic assumption of peer review, that there is a consensus on evaluation criteria, which has two main implications. First, the results offer an explanation for the disagreement effect. As there are two different consensuses on criteria, and as many scholars agree less with the criteria (i.e. those in the consensus-close and consensus-far classes), scholars are likely to use different criteria and weight criteria differently in grant peer review, which leads to different judgements. Naturally, this is a tentative and partial explanation, as the causality of the explanation has yet to be tested and there are other factors and mechanisms that can cause the disagreement effect (Brezis and Birukou, 2020; Lee, 2013; Lee, 2015; Sattler et al., 2015; Seeber et al., 2021). Criteria, however, might be an important factor in explaining the disagreement effect as Erosheva et al. (2020) found that scores of individual criteria 'play a major role in describing variability [of overall assessment scores in grant peer review], although they are not able to explain it fully' (p. 7).

Second, the results predict conservatism in peer review. This can be explained as follows. Research on group decision-making has shown that the more task-relevant preferences and cognitions are shared among members of a group, the greater their influence on the group's decision processes and outcomes (Tindale et al., 2003; Tindale and Winget, 2019). This implies that the criteria contained in both the core and broad consensus should exert the strongest influence in grant peer review. Specifically, this means that criteria related to knowledge gaps, feasibility, rigour, comprehensibility and argumentation, academic relevance, as well as to the competence and experience of the applicant should have a stronger influence than other criteria, such as originality and other contribution-related criteria. This is consistent with the notion of conservatism, which holds that innovative and ground-breaking research is undervalued in peer review (Lee et al., 2013), while other aspects, such as methodology (Lee, 2015) and feasibility (Langfeldt et al., 2020), are overweighted.

We proposed several things that a consensus on criteria could enable or promote. Specifically, we proposed that (a) it contributes to the validity, consistency, and efficiency of peer review, which increases its legitimacy, (b) it provides a common conceptual and semantic ground that enables and facilitates the interaction, communication, and information processing of peers, (c) it mitigates face-threatening acts occurring in evaluative contexts, and (d) it provides a way to determine the degree of peerness of a researcher. The results of this study suggest that review criteria are not sufficiently shared and weighted too differently to enable (a), (b), and (c). Hence, the legitimacy of peer review, common conceptual and semantic grounds, and face-threat mitigation have to be achieved by means other than a pre-existing consensus on criteria,[7] or a shared understanding of criteria has to be created during the review process, which then would facilitate (a), (b), and (c). In contrast, the two consensuses identified in this study could enable (d) the determination of the degree of peerness. Specifically, the degree to which a reviewer approves and applies the criteria of one of the two consensuses indicates her/his degree of membership in a scholarly community, how legitimate she/he is to make a judgement as a peer, and to what extent this judgement represents the judgement of a scholarly community.

The covariate analysis indicated that age and having tenure increases from the consensus-far to the consensus-close to the consensus classes. It also indicated that the percentage of scholars who have experience as grant reviewer is highest in the consensus classes. These results are in line with Gallo et al. (2018), who found that older reviewers rated evaluation criteria used at their last panel meeting as more appropriate than did younger reviewers. This suggests that the more academic experience scholars accumulate, the more their understanding of review criteria conforms to the consensus. These results could reflect the process of socialization, a key mechanism of social continuity and reproduction (Guhin, Calarco and Miller-Idriss, 2020): while senior and tenured scholars embody the norms (i.e. the two consensuses), younger scholars are learning, adapting to, and internalizing these norms.[8] Although the results of the covariate analysis for gender were less clear than those for age and having tenure, they suggest that (young and untenured) men are less conscious of criteria norms and that women might prefer a broader set of criteria (including contribution-related criteria) than men. Furthermore, the covariate analysis revealed that research field clearly differentiates between the two consensus-close classes. There are considerably more scholars from the field of languages and literatures in the class equivalent to the broad consensus (50%) than in the class equivalent to the core consensus (31%), and vice versa for the field of law (14% and 27%, respectively). Scholars from the field of history and cultural sciences are equally distributed in both consensus-close classes. These field differences are generally consistent with the literature, which finds that evaluation criteria differ between disciplines (e.g. Guetzkow, Lamont and Mallard, 2004; Hamann and Beljean, 2017; Hammarfelt, 2020; Hug et al., 2013; Langfeldt, Reymert and Aksnes, 2020; Reymert, Jungblut and Borlaug, 2020). The field differences between the consensus-close classes would offer a convenient explanation for the difference between the core and broad consensus, as the rating patterns of the consensus-close classes are almost identical to those of the consensus classes. However, we found no field differences between the two consensus-classes. These contradictory results might have methodological reasons. The three research fields might be too broad or the criteria included in the survey might be too general to detect differences reliably. Lastly, the covariate analysis indicated that neither the respondents' orientation of research (national, international, both) nor whether respondents rated criteria for an application of a doctoral student or a postdoctoral researcher predicted class membership. This suggests that the criteria have a wide scope and are applicable in different grant funding contexts. Overall, the covariate analysis was inconclusive in terms of factors distinguishing the core consensus from the broad consensus. We hypothesize that the two consensuses reflect different views on the predictability of the research process and its outcome. While the core consensus might reflect that some scholars acknowledge the uncertainty, serendipity, and unpredictability of research and thus prefer to ignore contribution-related criteria in grant peer review, the broad consensus might reflect that other scholars are more confident of the research process and hence factor expected outcomes into their assessment.

The results of this study have several implications for grant review practice. First, funding agencies as well as research on peer review have largely ignored the content validity of peer review (Hug and Aeschbach, 2020). The survey data collected and analysed in this study represent one element of content validation (Haynes, Richard and Kubany, 1995). Funding agencies may thus exploit the results presented here, particularly the criteria of the core and broad consensus, to design content-valid review procedures for the humanities. Additionally, the criteria included in the core consensus might be a suitable starting point for developing content-valid criteria in fields other than the humanities, since these criteria seem to be generalizable and not restricted to the humanities. Second, the fact that there are two criteria norms and that scholars have internalized them to varying degrees could be addressed on the side of the reviewers as well as on the side of the applicants. For example, funding agencies could make reviewers aware of the two norms, they could instruct and train reviewers to use the same criteria, or they could give reviewers the opportunity to explain and discuss their preferred criteria in order to promote a shared understanding of criteria and their importance. More importantly, younger and untenured applicants should be informed about the two criteria norms, as they seem to be less conscious of the norms and as receiving grants is vital to stay in academia and to secure a tenured position (Christian et al., 2021; Schröder, Lutter and Habicht, 2021). Younger researchers could also be more included in review panels to promote a shared understanding of criteria between younger and older scholars. Third, our study clearly showed that societal relevance is not important to humanities scholars in grant peer review. This is in stark contrast to research policy, where the notion of societal impact has gained significant traction in recent years, also with regard to the humanities (Pedersen, Grønvad and Hvidtfeldt, 2020). As funding agencies mediate between the research domain and the policy domain (Guston, 2000; Langfeldt et al., 2020; Reinhart, 2012), serious tensions regarding societal impact are to be expected in grant peer review if societal impact will be further emphasized. Humanities scholars and agencies may thus explore approaches to mitigate or avoid these looming tensions. Lastly, letters of recommendation can be safely removed from grant review procedures, as respondents have unanimously rejected this criterion.

We recognize three main limitations in our study. First, as the basic assumption has a broad range of meaning that can vary along at least four questions (who is considered a peer? does the assumption refer to criteria that peers prefer or apply? how is consensus defined? do criteria have to be shared or, more broadly, notions of quality, merit, excellence, or good science?) and as we have examined one particular version of the assumption, it is unclear whether our findings are valid for other versions of the assumption. Second, we have focused on grant peer review of doctoral and postdoctoral fellowships in the humanities in Switzerland. It remains to be investigated whether our findings can be generalized to other funding schemes, research fields, countries, and to other types of peer review. Third, the results might be distorted by response sets, that is, 'any tendency causing a person consistently to give different responses to test items than he would when the same content is presented in a different form' (Crohnbach, 1946, p. 476), such as satisficing, acquiescence or extreme response. Note, however, that according to survey research these behaviours have a social connotation and do not always represent bias but can be linked to personality traits (Couch and Keniston, 1960; Knowles & Nathan, 1997). Especially, in the context of this study, such response behaviour might reflect actual social behaviour in review situations. Therefore, further research might investigate whether such response styles represent bias (acquiescence or satisficing) or rather substantial patterns (convictions and social behaviour) using cognitive interviews and measurements of response styles.

As this is an exploratory study, future studies may translate the findings presented here into hypotheses and test them empirically. Future studies may also investigate reasons why the two criteria consensuses differ, as we found no clear evidence of whether the consensuses are related to research fields, gender, or other factors. Finally, future studies may explicate and examine other basic assumptions of peer review, which could pave the way for building theories of peer review.

**Notes**

1. For the emergence of modern peer review, see Baldwin (2017, 2018, 2020) and Moxham and Fyfe (2018).
2. Low inter-rater reliability is observed when reviewers or panellists individually assess research, typically at the beginning or in the middle of the peer review process. The process as a whole, however, (mostly) resolves disagreement and produces a consensual decision, such as 'reject/accept' or 'fund/do not fund' (Lamont, 2009; Reinhart, 2012).
3. Recently, Erosheva et al. (2021) argued that the way inter-rater reliability is calculated can strongly influence reliability coefficients. They therefore recommended against using inter-rater reliability to assess the quality of peer review.
4. For definitions of legitimacy, see Tyler (2006) and Johnson, Dowd and Ridgeway (2006).
5. Our presentation at the ISSI 2017 Conference was based on similar research questions, preliminary data, and a different statistical approach (Ochsner, Hug and Daniel, 2017).
6. Note that the findings of Tindale et al. (2003) apply to the entire peer review process, including individual reviews and panels, as research on group decision-making defines its scope broadly and studies groups without interaction and decision control as well as groups with full interaction and decision control (Tindale and Winget, 2019).
7. Lamont (2009) discusses such means under the heading of 'technology of peer review' and 'customary rules of deliberation' and Reinhart (2012) under the heading of '*organisationale*

*Struktur des Entscheidungsverfahrens* [organizational structure of decision-making procedures]'.

8. There are other possible explanations for these results, such as social distinction (i.e. experienced researchers and reviewers agree on criteria over time to preserve their power; see e.g. Bourdieu (1976) or selection bias (i.e. scholars who endorse criteria considered relevant by experienced researchers are more likely to be assigned a senior or gatekeeping position than those who adhere to other criteria; see e.g. van den Brink and Benschop (2012). Naturally, reviewers can also learn from, or simply internalize, the criteria that funding agencies prescribe.

## Acknowledgments

We thank Klaus Jonas, Martin Reinhart, Rüdiger Mutz, and the reviewers for comments on the manuscript. This study was supported by Swissuniversities, SUK Program P-3 'Performances de la recherche en sciences humaines et sociales'.

# Do peers share the same criteria for assessing grant applications?

Sven E. Hug & Michael Ochsner

# Supplementary material

# Table of contents

# Part A: Methods

Table 1. Evaluation criteria rated in the survey. The labels of the six-point scale were (1) strongly disagree, (2) disagree, (3) slightly disagree, (4) slightly agree, (5) agree, and (6) strongly agree.

| Variable name | Group | Item name | Item description |
| --- | --- | --- | --- |
| *An application is assessed appropriately, if the assessment considers whether ...* | | | |
| Eigenst | | Independence | the project is the applicant's own research project (is not dominated by others, e.g. supervisor, interest groups; does not form part of a bigger research programme designed by others). |
| Orig1 | Originality | Gaps | the application identifies gaps in existing knowledge. |
| Orig2 | Originality | New approach | the project uses a new approach to a research topic or to data (e.g. new research question, reading, perspective, theory, method). |
| Orig3 | Originality | New topic | the project introduces a new research topic. |
| Orig4 | Originality | Innovative data | the project is innovative with regard to the data (e.g. capturing/discovering new data; identifying little known, neglected or not yet discussed data; putting known data together in a new way). |
| Orig5 | Originality | New paradigm | the project generates a new paradigm (e.g. new way of thinking, theory, method; opens up a new area of research). |
| Orig6 | Originality | New findings | the project generates new findings within existing paradigms (new interpretations within existing theories, methods, research areas, established research questions, or schools of thought). |
| Real1 | Feasibility | Timetable | the project timetable is realistic. |
| Real2 | Feasibility | Resources | the project can be implemented with the planned resources (e.g. personnel, infrastructure, funding). |
| Rigour1 | Rigour | State of res. | the current state of research is expressed appropriately. |
| Rigour2 | Rigour | Method choice | the choice of method is appropriate (e.g. suitability of the method; reflection on choice of method). |
| Rigour3 | Rigour | Res. process | the suggested research process is appropriate (e.g. planning, organisation, implementation of the project). |

| Variable name | Group | Item name | Item description |
| --- | --- | --- | --- |
| Rigour4 | Rigour | Argumentation | the argumentation is stringent (e.g. clear and logical development of the research question and hypotheses; logically coordinated steps of the investigation; coherent argumentation). |
| Rigour5 | Rigour | Comprehensibility | the project is presented in a comprehensible manner (e.g. understandable language; clear structure). |
| Rel1 | Relevance | Acad. relevance | the expected findings are relevant for academia. |
| Rel2 | Relevance | Soc. relevance | the expected findings are relevant for society (i.e. for groups and stakeholders outside academia). |
| KultErb | | Cult. heritage | the project contributes to fostering cultural heritage (e.g. documentation, preservation, or keeping alive aspects of the past; renews the understanding and interpretation of aspects of the past). |
| Komplex | | Complexity | the project makes complexity visible (e.g. works against the trend towards reducing complexity; shows the ambivalence, ambiguity and complexity of the research topic; takes the many facets of a question seriously). |
| Vielfalt | | Variety | the project contributes to the variety of research (e.g. with regard to research topics, approaches, theories, methods, data; it takes great risks; it is outside the mainstream). |
| *An application is assessed appropriately, if...* | | | |
| P_CV | Applicant | CV | the person's CV is taken into consideration (e.g. acquired competences, research experience, third-party funding acquired, work performed in relation to academic age, awards, international experience or quality of the institute/chair to which the applicant is/was affiliated). |
| P_Dipl | Applicant | Diploma | the applicant's degrees/qualifications are taken into account (e.g. quality of the degrees acquired, marks obtained for qualifications, quality of the institute/chair at which the qualifications were obtained). |
| P_PubList | Applicant | Publications | the applicant's publication list is taken into consideration (e.g. scope, quality). |
| P_Letter | Applicant | Letter | the letters of recommendation for the applicant are taken into consideration (e.g. from doctoral advisors, previous employers). |

Table 2. Personal information collected prior to the survey.

| Variable name | Response options |
|---|---|
| Gender | female (w) <br> male (m) |
| Doc | application of a doctoral student (doc) <br> application of a postdoctoral researcher (postdoc) |
| Field | languages and literatures (Lang) <br> history and cultural sciences (HistCult) <br> law (Law) |

Table 3. Personal information collected in the survey.

| Variable name | Item content | Response options |
|---|---|---|
| Age | Year of birth | year |
| Tenured | Are you a tenured professor? (Ordinaria, Ordinarius) | yes <br> no |
| NatInternat | The orientation of my research is ... | mainly national. (nat) <br> mainly international. (intl) <br> equally national and international. (natintl) |
| GPR | I serve / have served as a reviewer of grant proposals. | yes (Grant) <br> no (NotGrant) |

**Evaluation criteria used by the Swiss National Science Foundation at the time of the survey (English version)**

Main criteria defined in the funding regulations of the Swiss National Science Foundation:
− Scientific value and relevance of the project
− Originality of research objectives
− Adequacy of methodical approach
− Feasibility of the project
− Scientific track record of the applicants
− Applicants' expertise in relation to the project

Criteria defined in the regulations for Doc.CH grants (funding of PhD students):
− Quality, originality, topicality, and feasibility of the dissertation project

- Applicant's contribution towards the topic and concept of the dissertation project
- Applicant's scientific track record
- Likelihood of the dissertation being successfully completed
- Applicant's aptitude for a scientific career and the prospects for such a career in Switzerland
- Applicant's mobility as regards the envisaged place of work
- Quality of the research location envisaged for the dissertation, in particular the working conditions as well as the possibilities for professional supervision and further education
- Qualifications of the dissertation supervisor

Criteria defined in the regulations for Ambizione grants (funding of postdocs):
- Quality, originality, relevance, and independence of the research project for which the Ambizione grant is envisaged
- Scientific autonomy of the applicant at the host institute
- Applicant's scientific track record, in particular research work and resultant publications
- Personal suitability of the applicant for a high-level career in academic/clinical research
- Proof of mobility of the applicant prior to the submission of the proposal and with regard to the choice of workplace
- Potential for integration in the Swiss scientific community

**R code for estimating Latent Class Trees and predicting class membership**

```
## Install package
install.packages("devtools")
library(devtools)
install_github("MattisvdBergh/LCT")
library(LCTpackage)

## Prepare analysis
Dat <- read.csv("data.csv", header=T, sep=",")
LG <- "C:/PATH/LatentGOLD5.1/lg51.exe"
itemNames <- c("Eigenst","Orig1","Orig2","Orig3","Orig4","Orig5","Orig6","Real1",
               "Real2","Rigour1","Rigour2","Rigour3","Rigour4","Rigour5","Rel1",
               "Rel2","KultErb","Komplex","Vielfalt","P_CV","P_Dipl","P_PubList",
               "P_Letter")
Dat[itemNames] <- sapply(Dat[,itemNames],function(x){as.factor(x)})

## Estimate Latent Class Tree
Results.SC3 <- LCT(Dataset=Dat,
         LG=LG,
         maxClassSplit1=4,
         resultsName="_LCT",
         itemNames=itemNames,
         stopCriterium="BIC",
         nKeepVariables=7,
         namesKeepVariables= c("Gender","Doc","Tenured","NatInternat","Field",
```

```
                                  "Age","GPR")
              )

## Predict class membership
explTree.SC3 <- exploreTree(resTree=Results.SC3,
                  dirTreeResults="C:/PATH/Results_LCT",
                  ResultsFolder="_explTree",
                  analysis="covariates",
                Covariates=c("Gender","Doc","Tenured","NatInternat","Field","Age",
                                "GPR"),
                  sizeMlevels = c(2,2,2,3,3,1,2),
                  mLevels = c("Nominal","Nominal","Nominal","Nominal","Nominal",
                                "continuous","Nominal"),
                  method="ml")
```

# Part B: Results

Table 4. Descriptive statistics of the personal characteristics (population and samples).

| Characteristics | N = 2,609 | n = 866 | n = 760 |
|---|---|---|---|
| Gender | | | |
|     Men | 63% | 59% | 59% |
|     Women | 37% | 41% | 41% |
| Criteria for applications of | | | |
|     Doctoral students | 51% | 53% | 53% |
|     Postdoctoral researchers | 49% | 47% | 47% |
| Field | | | |
|     Languages and literatures | 37% | 41% | 40% |
|     History and cultural sciences | 35% | 41% | 41% |
|     Law | 28% | 18% | 18% |
| Age | | | |
|     Mean | – | 48.38 | 47.98 |
|     Median | | 47.50 | 47.00 |
| Tenured | | | |
|     Yes | – | 39% | 38% |
|     No | | 61% | 62% |
| Orientation of research | | | |
|     Mainly national | – | 14% | 14% |
|     Mainly international | | 41% | 40% |
|     Equally national and international | | 45% | 45% |
| Served as reviewer of grant proposals | | | |
|     Yes | – | 56% | 55% |
|     No | | 44% | 45% |

## Sample n = 866

### Latent Class Tree

Table 5. Fit statistics and their relative improvement (n = 866). Number of classes per model, log-likelihood, number of parameters, BIC, AIC, and relative improvement of fit of the log-likelihood, BIC, and AIC.

| # classes | logL | P | BIC | AIC | $RI_{logL}$ | $RI_{BIC}$ | $RI_{AIC}$ |
|---|---|---|---|---|---|---|---|
| 1 | -25677 | 115 | 52132 | 51584 | | | |
| 2 | -24582 | 139 | 50104 | 49442 | 1.000 | 1.000 | 1.000 |
| 3 | -24269 | 163 | 49640 | 48864 | 0.286 | 0.229 | 0.270 |
| 4 | -24021 | 187 | 49308 | 48417 | 0.226 | 0.164 | 0.209 |
| 5 | -23871 | 211 | 49168 | 48163 | 0.138 | 0.069 | 0.118 |
| 6 | -23749 | 235 | 49088 | 47969 | 0.111 | 0.040 | 0.091 |
| 7 | -23650 | 259 | 49051 | 47818 | 0.091 | 0.018 | 0.071 |
| 8 | -23571 | 283 | 49057 | 47709 | 0.072 | -0.003 | 0.051 |
| 9 | -23472 | 307 | 49021 | 47558 | 0.090 | 0.018 | 0.070 |
| 10 | -23443 | 331 | 49124 | 47548 | 0.027 | -0.051 | 0.005 |

Table 6. Wald statistics of the paired comparisons of the main classes LC 1, 2, 3, and 4.

| Models for indicators | | | Wald | p |
|---|---|---|---|---|
| Eigenst | | | | |
| LC | 1 | 2 | 0.314 | 0.58 |
| LC | 1 | 3 | 19.168 | 1.20E-05 |
| LC | 1 | 4 | 1.552 | 0.21 |
| LC | 2 | 3 | 17.759 | 2.50E-05 |
| LC | 2 | 4 | 0.525 | 0.47 |
| LC | 3 | 4 | 20.439 | 6.20E-06 |
| Orig1 | | | | |
| LC | 1 | 2 | 9.432 | 0.0021 |
| LC | 1 | 3 | 52.281 | 4.80E-13 |
| LC | 1 | 4 | 24.309 | 8.20E-07 |
| LC | 2 | 3 | 19.093 | 1.20E-05 |
| LC | 2 | 4 | 37.442 | 9.40E-10 |
| LC | 3 | 4 | 82.792 | 9.10E-20 |
| Orig2 | | | | |
| LC | 1 | 2 | 0.479 | 0.49 |
| LC | 1 | 3 | 55.573 | 9.00E-14 |
| LC | 1 | 4 | 11.270 | 0.00079 |
| LC | 2 | 3 | 60.241 | 8.40E-15 |
| LC | 2 | 4 | 5.955 | 0.015 |
| LC | 3 | 4 | 74.226 | 7.00E-18 |
| Orig3 | | | | |
| LC | 1 | 2 | 23.357 | 1.30E-06 |
| LC | 1 | 3 | 42.206 | 8.20E-11 |
| LC | 1 | 4 | 7.828 | 0.0052 |
| LC | 2 | 3 | 76.859 | 1.80E-18 |

| Models for indicators | | | Wald | p |
|---|---|---|---|---|
| LC | 2 | 4 | 2.075 | 0.15 |
| LC | 3 | 4 | 57.967 | 2.70E-14 |
| Orig4 | | | | |
| LC | 1 | 2 | 3.995 | 0.046 |
| LC | 1 | 3 | 75.068 | 4.50E-18 |
| LC | 1 | 4 | 13.793 | 0.0002 |
| LC | 2 | 3 | 91.219 | 1.30E-21 |
| LC | 2 | 4 | 3.139 | 0.076 |
| LC | 3 | 4 | 98.414 | 3.40E-23 |
| Orig5 | | | | |
| LC | 1 | 2 | 31.355 | 2.10E-08 |
| LC | 1 | 3 | 32.396 | 1.30E-08 |
| LC | 1 | 4 | 14.740 | 0.00012 |
| LC | 2 | 3 | 79.337 | 5.20E-19 |
| LC | 2 | 4 | 0.604 | 0.44 |
| LC | 3 | 4 | 58.951 | 1.60E-14 |
| Orig6 | | | | |
| LC | 1 | 2 | 0.316 | 0.57 |
| LC | 1 | 3 | 60.929 | 5.90E-15 |
| LC | 1 | 4 | 21.360 | 3.80E-06 |
| LC | 2 | 3 | 61.731 | 3.90E-15 |
| LC | 2 | 4 | 14.320 | 0.00015 |
| LC | 3 | 4 | 86.744 | 1.20E-20 |
| Real1 | | | | |
| LC | 1 | 2 | 44.596 | 2.40E-11 |
| LC | 1 | 3 | 57.372 | 3.60E-14 |
| LC | 1 | 4 | 40.699 | 1.80E-10 |
| LC | 2 | 3 | 0.248 | 0.62 |
| LC | 2 | 4 | 82.444 | 1.10E-19 |
| LC | 3 | 4 | 106.397 | 6.00E-25 |
| Real2 | | | | |
| LC | 1 | 2 | 66.232 | 4.00E-16 |
| LC | 1 | 3 | 67.785 | 1.80E-16 |
| LC | 1 | 4 | 37.867 | 7.60E-10 |
| LC | 2 | 3 | 0.328 | 0.57 |
| LC | 2 | 4 | 107.444 | 3.60E-25 |
| LC | 3 | 4 | 113.071 | 2.10E-26 |
| Rigour1 | | | | |
| LC | 1 | 2 | 75.443 | 3.80E-18 |
| LC | 1 | 3 | 79.597 | 4.60E-19 |
| LC | 1 | 4 | 39.476 | 3.30E-10 |
| LC | 2 | 3 | 0.008 | 0.93 |
| LC | 2 | 4 | 114.876 | 8.40E-27 |
| LC | 3 | 4 | 125.600 | 3.80E-29 |
| Rigour2 | | | | |
| LC | 1 | 2 | 64.223 | 1.10E-15 |
| LC | 1 | 3 | 89.622 | 2.90E-21 |
| LC | 1 | 4 | 36.666 | 1.40E-09 |
| LC | 2 | 3 | 0.518 | 0.47 |
| LC | 2 | 4 | 95.012 | 1.90E-22 |
| LC | 3 | 4 | 131.820 | 1.60E-30 |
| Rigour3 | | | | |
| LC | 1 | 2 | 65.524 | 5.70E-16 |
| LC | 1 | 3 | 114.483 | 1.00E-26 |
| LC | 1 | 4 | 52.210 | 5.00E-13 |
| LC | 2 | 3 | 0.354 | 0.55 |

| Models for indicators | | | Wald | p |
|---|---|---|---|---|
| LC | 2 | 4 | 105.114 | 1.20E-24 |
| LC | 3 | 4 | 165.979 | 5.60E-38 |
| Rigour4 | | | | |
| LC | 1 | 2 | 38.347 | 5.90E-10 |
| LC | 1 | 3 | 51.631 | 6.70E-13 |
| LC | 1 | 4 | 27.033 | 2.00E-07 |
| LC | 2 | 3 | 0.963 | 0.33 |
| LC | 2 | 4 | 60.818 | 6.30E-15 |
| LC | 3 | 4 | 86.024 | 1.80E-20 |
| Rigour5 | | | | |
| LC | 1 | 2 | 45.673 | 1.40E-11 |
| LC | 1 | 3 | 47.697 | 5.00E-12 |
| LC | 1 | 4 | 20.626 | 5.60E-06 |
| LC | 2 | 3 | 0.038 | 0.84 |
| LC | 2 | 4 | 72.305 | 1.80E-17 |
| LC | 3 | 4 | 75.413 | 3.80E-18 |
| Rel1 | | | | |
| LC | 1 | 2 | 13.084 | 0.0003 |
| LC | 1 | 3 | 68.905 | 1.00E-16 |
| LC | 1 | 4 | 27.804 | 1.30E-07 |
| LC | 2 | 3 | 31.367 | 2.10E-08 |
| LC | 2 | 4 | 44.669 | 2.30E-11 |
| LC | 3 | 4 | 99.640 | 1.80E-23 |
| Rel2 | | | | |
| LC | 1 | 2 | 2.589 | 0.11 |
| LC | 1 | 3 | 31.977 | 1.60E-08 |
| LC | 1 | 4 | 22.228 | 2.40E-06 |
| LC | 2 | 3 | 35.242 | 2.90E-09 |
| LC | 2 | 4 | 10.118 | 0.0015 |
| LC | 3 | 4 | 62.127 | 3.20E-15 |
| KultErb | | | | |
| LC | 1 | 2 | 2.467 | 0.12 |
| LC | 1 | 3 | 28.099 | 1.20E-07 |
| LC | 1 | 4 | 13.431 | 0.00025 |
| LC | 2 | 3 | 35.183 | 3.00E-09 |
| LC | 2 | 4 | 4.432 | 0.035 |
| LC | 3 | 4 | 50.534 | 1.20E-12 |
| Komplex | | | | |
| LC | 1 | 2 | 0.082 | 0.77 |
| LC | 1 | 3 | 35.608 | 2.40E-09 |
| LC | 1 | 4 | 6.127 | 0.013 |
| LC | 2 | 3 | 24.500 | 7.40E-07 |
| LC | 2 | 4 | 6.051 | 0.014 |
| LC | 3 | 4 | 44.787 | 2.20E-11 |
| Vielfalt | | | | |
| LC | 1 | 2 | 0.163 | 0.69 |
| LC | 1 | 3 | 31.746 | 1.80E-08 |
| LC | 1 | 4 | 3.701 | 0.054 |
| LC | 2 | 3 | 28.176 | 1.10E-07 |
| LC | 2 | 4 | 2.285 | 0.13 |
| LC | 3 | 4 | 36.161 | 1.80E-09 |
| P_CV | | | | |
| LC | 1 | 2 | 20.757 | 5.20E-06 |
| LC | 1 | 3 | 35.586 | 2.40E-09 |
| LC | 1 | 4 | 50.069 | 1.50E-12 |
| LC | 2 | 3 | 0.186 | 0.67 |

| Models for indicators | | | Wald | p |
|---|---|---|---|---|
| | LC | 2 4 | 64.601 | 9.20E-16 |
| | LC | 3 4 | 91.107 | 1.40E-21 |
| P_Dipl | | | | |
| | LC | 1 2 | 15.449 | 8.50E-05 |
| | LC | 1 3 | 35.655 | 2.40E-09 |
| | LC | 1 4 | 49.933 | 1.60E-12 |
| | LC | 2 3 | 2.228 | 0.14 |
| | LC | 2 4 | 63.494 | 1.60E-15 |
| | LC | 3 4 | 90.629 | 1.70E-21 |
| P_PubList | | | | |
| | LC | 1 2 | 15.630 | 7.70E-05 |
| | LC | 1 3 | 34.162 | 5.10E-09 |
| | LC | 1 4 | 36.294 | 1.70E-09 |
| | LC | 2 3 | 1.496 | 0.22 |
| | LC | 2 4 | 56.886 | 4.60E-14 |
| | LC | 3 4 | 79.839 | 4.10E-19 |
| P_Letter | | | | |
| | LC | 1 2 | 3.198 | 0.074 |
| | LC | 1 3 | 17.691 | 2.60E-05 |
| | LC | 1 4 | 52.532 | 4.20E-13 |
| | LC | 2 3 | 3.572 | 0.059 |
| | LC | 2 4 | 55.232 | 1.10E-13 |
| | LC | 3 4 | 79.331 | 5.30E-19 |

Table 7. Wald statistics of the paired comparisons of the subclasses LC 1.1 and 1.2.

| Models for indicators | | | Wald | p |
|---|---|---|---|---|
| Eigenst | | | | |
| | LC | 1.1 1.2 | 1.156 | 0.28 |
| Orig1 | | | | |
| | LC | 1.1 1.2 | 0.456 | 0.5 |
| Orig2 | | | | |
| | LC | 1.1 1.2 | 23.633 | 1.20E-06 |
| Orig3 | | | | |
| | LC | 1.1 1.2 | 7.955 | 0.0048 |
| Orig4 | | | | |
| | LC | 1.1 1.2 | 23.106 | 1.50E-06 |
| Orig5 | | | | |
| | LC | 1.1 1.2 | 18.416 | 1.80E-05 |
| Orig6 | | | | |
| | LC | 1.1 1.2 | 10.431 | 0.0012 |
| Real1 | | | | |
| | LC | 1.1 1.2 | 0.698 | 0.4 |
| Real2 | | | | |
| | LC | 1.1 1.2 | 0.206 | 0.65 |
| Rigour1 | | | | |
| | LC | 1.1 1.2 | 0.000 | 0.99 |
| Rigour2 | | | | |
| | LC | 1.1 1.2 | 1.571 | 0.21 |

| Models for indicators | | | Wald | p |
|---|---|---|---|---|
| Rigour3 | | | | |
| LC | 1.1 | 1.2 | 1.700 | 0.19 |
| Rigour4 | | | | |
| LC | 1.1 | 1.2 | 0.338 | 0.56 |
| Rigour5 | | | | |
| LC | 1.1 | 1.2 | 0.020 | 0.89 |
| Rel1 | | | | |
| LC | 1.1 | 1.2 | 0.018 | 0.89 |
| Rel2 | | | | |
| LC | 1.1 | 1.2 | 0.528 | 0.47 |
| KultErb | | | | |
| LC | 1.1 | 1.2 | 9.669 | 0.0019 |
| Komplex | | | | |
| LC | 1.1 | 1.2 | 5.732 | 0.017 |
| Vielfalt | | | | |
| LC | 1.1 | 1.2 | 10.628 | 0.0011 |
| P_CV | | | | |
| LC | 1.1 | 1.2 | 1.572 | 0.21 |
| P_Dipl | | | | |
| LC | 1.1 | 1.2 | 1.705 | 0.19 |
| P_PubList | | | | |
| LC | 1.1 | 1.2 | 1.734 | 0.19 |
| P_Letter | | | | |
| LC | 1.1 | 1.2 | 1.392 | 0.24 |

Table 8. Wald statistics of the paired comparisons of the subclasses LC 3.1 and 3.2.

| Models for indicators | | | Wald | p |
|---|---|---|---|---|
| Eigenst | | | | |
| LC | 3.1 | 3.2 | 0.099 | 0.75 |
| Orig1 | | | | |
| LC | 3.1 | 3.2 | 0.254 | 0.61 |
| Orig2 | | | | |
| LC | 3.1 | 3.2 | 0.019 | 0.89 |
| Orig3 | | | | |
| LC | 3.1 | 3.2 | 0.821 | 0.36 |
| Orig4 | | | | |
| LC | 3.1 | 3.2 | 0.987 | 0.32 |
| Orig5 | | | | |
| LC | 3.1 | 3.2 | 1.678 | 0.2 |
| Orig6 | | | | |
| LC | 3.1 | 3.2 | 0.972 | 0.32 |
| Real1 | | | | |
| LC | 3.1 | 3.2 | 139.623 | 3.2E-32 |
| Real2 | | | | |
| LC | 3.1 | 3.2 | 39.856 | 2.7E-10 |
| Rigour1 | | | | |
| LC | 3.1 | 3.2 | 5.789 | 0.016 |

| Models for indicators | | | Wald | p |
|---|---|---|---|---|
| Rigour2 | | | | |
| LC | 3.1 | 3.2 | 1.846 | 0.17 |
| Rigour3 | | | | |
| LC | 3.1 | 3.2 | 8.714 | 0.0032 |
| Rigour4 | | | | |
| LC | 3.1 | 3.2 | 5.011 | 0.025 |
| Rigour5 | | | | |
| LC | 3.1 | 3.2 | 4.975 | 0.026 |
| Rel1 | | | | |
| LC | 3.1 | 3.2 | 0.530 | 0.47 |
| Rel2 | | | | |
| LC | 3.1 | 3.2 | 3.722 | 0.054 |
| KultErb | | | | |
| LC | 3.1 | 3.2 | 2.473 | 0.12 |
| Komplex | | | | |
| LC | 3.1 | 3.2 | 2.352 | 0.13 |
| Vielfalt | | | | |
| LC | 3.1 | 3.2 | 0.746 | 0.39 |
| P_CV | | | | |
| LC | 3.1 | 3.2 | 2.933 | 0.087 |
| P_Dipl | | | | |
| LC | 3.1 | 3.2 | 5.894 | 0.015 |
| P_PubList | | | | |
| LC | 3.1 | 3.2 | 1.146 | 0.28 |
| P_Letter | | | | |
| LC | 3.1 | 3.2 | 5.274 | 0.022 |

## Sample n = 760

### Latent Class Tree

Table 9. Fit statistics and their relative improvement (n = 760). Number of classes per model, log-likelihood, number of parameters, BIC, AIC, and relative improvement of fit of the log-likelihood, BIC, and AIC.

| # classes | logL | P | BIC | AIC | $RI_{logL}$ | $RI_{BIC}$ | $RI_{AIC}$ |
|---|---|---|---|---|---|---|---|
| 1 | -22736 | 115 | 46235 | 45702 | | | |
| 2 | -21770 | 139 | 44462 | 43818 | 1.000 | 1.000 | 1.000 |
| 3 | -21495 | 163 | 44072 | 43317 | 0.284 | 0.220 | 0.266 |
| 4 | -21265 | 187 | 43770 | 42904 | 0.238 | 0.170 | 0.219 |
| 5 | -21115 | 211 | 43630 | 42652 | 0.155 | 0.079 | 0.134 |
| 6 | -21019 | 235 | 43598 | 42509 | 0.099 | 0.018 | 0.076 |
| 7 | -20906 | 259 | 43529 | 42329 | 0.118 | 0.039 | 0.095 |
| 8 | -20865 | 283 | 43607 | 42296 | 0.042 | -0.044 | 0.018 |
| 9 | -20749 | 307 | 43535 | 42112 | 0.120 | 0.041 | 0.097 |
| 10 | -20693 | 331 | 43582 | 42048 | 0.058 | -0.027 | 0.034 |

Figure 1. Latent class tree (n = 760). Class size is indicated in percent of the sample size.

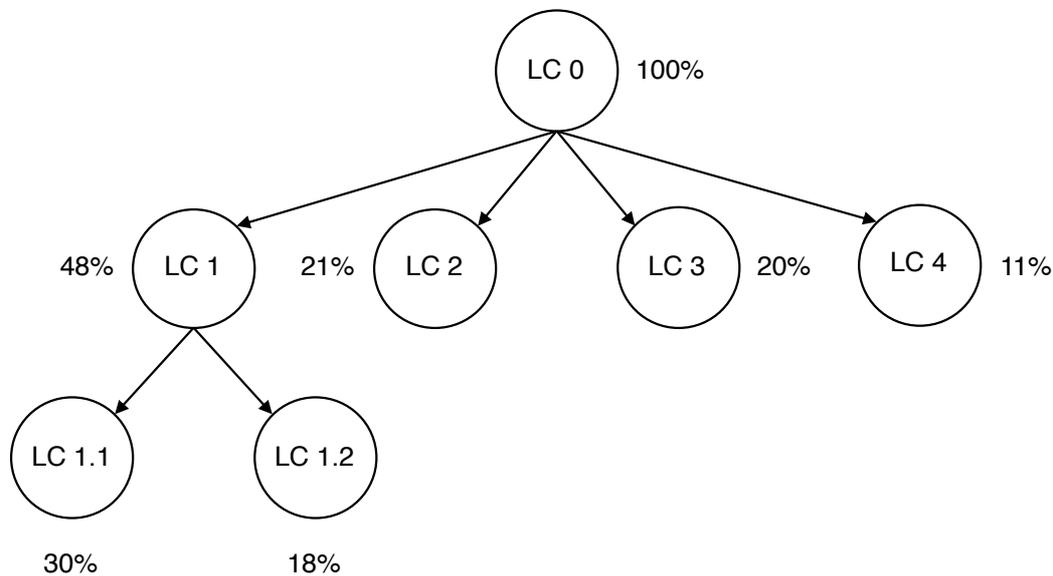

The expansion of the tree from one to four classes yields a very large improvement of the BIC ($\Delta BIC = 2'465$), while splitting LC 1 ($\Delta BIC = 78$) yields a marginal improvement.

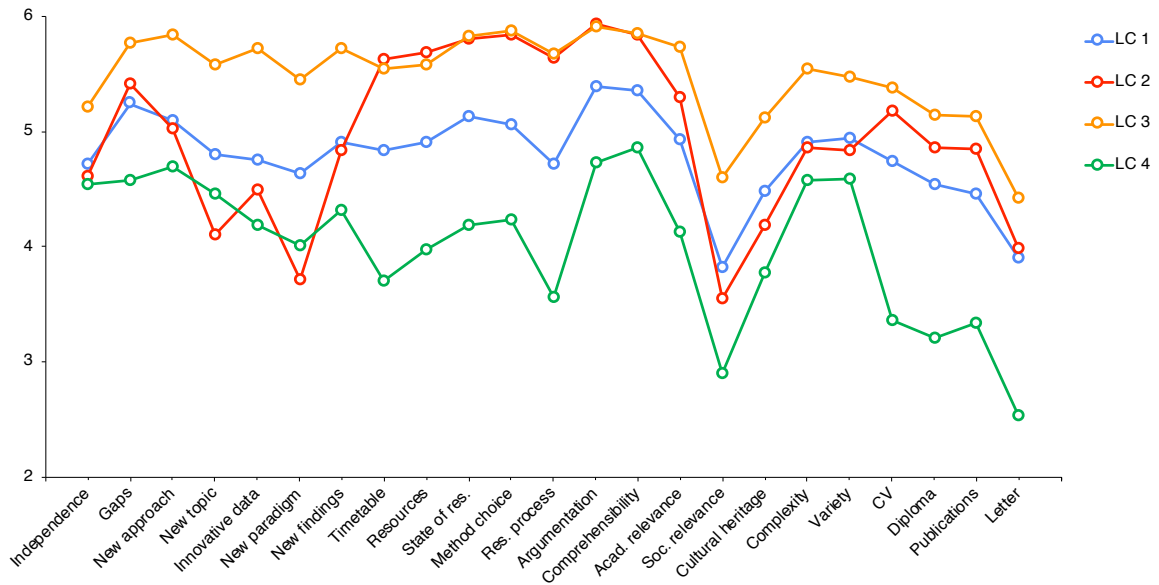

Figure 2. Average rating of the criteria by main classes of the latent class tree (n = 760) on a scale from 1 (strongly disagree) to 6 (strongly agree).

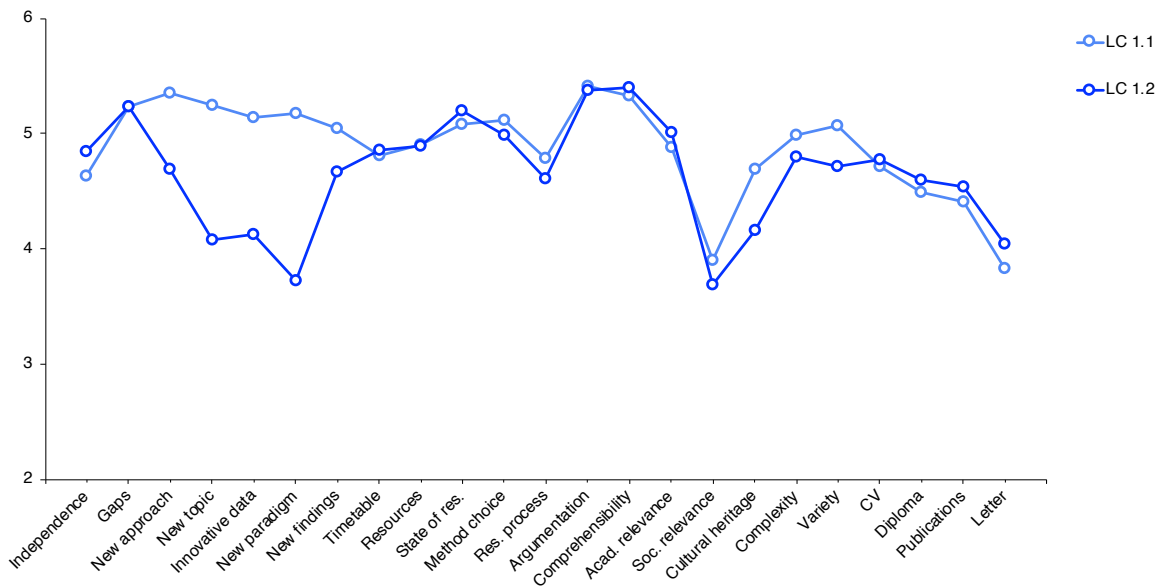

Figure 3. Average rating of the criteria by subclasses of the latent class tree (n = 760) on a scale from 1 (strongly disagree) to 6 (strongly agree).

Figure 4. Probability of rating a criterion as 5 (agree) or 6 (strongly agree) by classes of the final latent class tree (n = 760). The dashed line indicates a probability of 0.85.

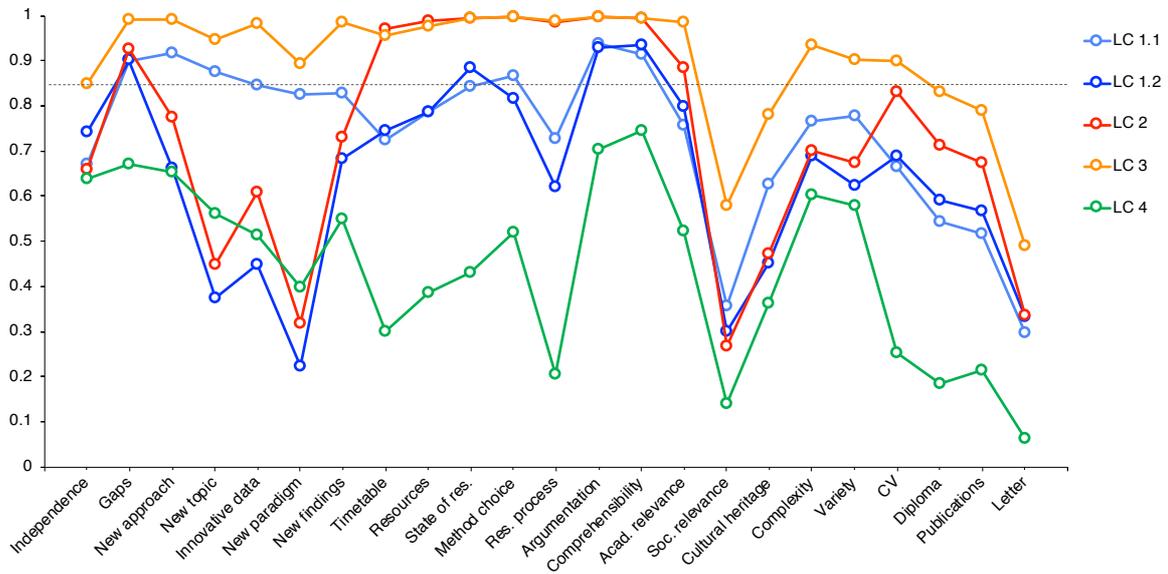

Table 10. Wald statistics of the paired comparisons of the main classes LC 1, 2, 3, and 4.

| Models for indicators | | | Wald | p |
|---|---|---|---|---|
| Eigenst | | | | |
| LC | 1 | 2 | 0.582 | 0.45 |
| LC | 1 | 3 | 13.846 | 0.0002 |
| LC | 1 | 4 | 1.012 | 0.31 |
| LC | 2 | 3 | 13.870 | 0.0002 |
| LC | 2 | 4 | 0.122 | 0.73 |
| LC | 3 | 4 | 14.284 | 0.00016 |
| Orig1 | | | | |
| LC | 1 | 2 | 3.644 | 0.056 |
| LC | 1 | 3 | 41.969 | 9.30E-11 |
| LC | 1 | 4 | 23.196 | 1.50E-06 |
| LC | 2 | 3 | 19.907 | 8.10E-06 |
| LC | 2 | 4 | 26.078 | 3.30E-07 |
| LC | 3 | 4 | 69.727 | 6.80E-17 |
| Orig2 | | | | |
| LC | 1 | 2 | 0.602 | 0.44 |
| LC | 1 | 3 | 49.927 | 1.60E-12 |
| LC | 1 | 4 | 10.563 | 0.0012 |
| LC | 2 | 3 | 52.218 | 5.00E-13 |
| LC | 2 | 4 | 4.926 | 0.026 |
| LC | 3 | 4 | 66.470 | 3.60E-16 |
| Orig3 | | | | |
| LC | 1 | 2 | 20.672 | 5.50E-06 |
| LC | 1 | 3 | 36.373 | 1.60E-09 |
| LC | 1 | 4 | 4.532 | 0.033 |
| LC | 2 | 3 | 65.861 | 4.80E-16 |

| Models for indicators | | | Wald | p |
|---|---|---|---|---|
| LC | 2 | 4 | 3.011 | 0.083 |
| LC | 3 | 4 | 45.554 | 1.50E-11 |
| Orig4 | | | | |
| LC | 1 | 2 | 3.107 | 0.078 |
| LC | 1 | 3 | 61.412 | 4.60E-15 |
| LC | 1 | 4 | 10.099 | 0.0015 |
| LC | 2 | 3 | 70.835 | 3.90E-17 |
| LC | 2 | 4 | 2.268 | 0.13 |
| LC | 3 | 4 | 77.985 | 1.00E-18 |
| Orig5 | | | | |
| LC | 1 | 2 | 29.597 | 5.30E-08 |
| LC | 1 | 3 | 31.602 | 1.90E-08 |
| LC | 1 | 4 | 10.069 | 0.0015 |
| LC | 2 | 3 | 79.032 | 6.10E-19 |
| LC | 2 | 4 | 1.815 | 0.18 |
| LC | 3 | 4 | 51.519 | 7.10E-13 |
| Orig6 | | | | |
| LC | 1 | 2 | 0.478 | 0.49 |
| LC | 1 | 3 | 50.928 | 9.60E-13 |
| LC | 1 | 4 | 16.330 | 5.30E-05 |
| LC | 2 | 3 | 51.100 | 8.80E-13 |
| LC | 2 | 4 | 9.748 | 0.0018 |
| LC | 3 | 4 | 70.501 | 4.60E-17 |
| Real1 | | | | |
| LC | 1 | 2 | 40.689 | 1.80E-10 |
| LC | 1 | 3 | 53.903 | 2.10E-13 |
| LC | 1 | 4 | 40.704 | 1.80E-10 |
| LC | 2 | 3 | 0.981 | 0.32 |
| LC | 2 | 4 | 76.563 | 2.10E-18 |
| LC | 3 | 4 | 102.008 | 5.50E-24 |
| Real2 | | | | |
| LC | 1 | 2 | 54.687 | 1.40E-13 |
| LC | 1 | 3 | 59.433 | 1.30E-14 |
| LC | 1 | 4 | 38.025 | 7.00E-10 |
| LC | 2 | 3 | 1.689 | 0.19 |
| LC | 2 | 4 | 93.395 | 4.30E-22 |
| LC | 3 | 4 | 104.333 | 1.70E-24 |
| Rigour1 | | | | |
| LC | 1 | 2 | 61.825 | 3.80E-15 |
| LC | 1 | 3 | 68.588 | 1.20E-16 |
| LC | 1 | 4 | 41.253 | 1.30E-10 |
| LC | 2 | 3 | 0.123 | 0.73 |
| LC | 2 | 4 | 102.209 | 5.00E-24 |
| LC | 3 | 4 | 114.567 | 9.80E-27 |
| Rigour2 | | | | |
| LC | 1 | 2 | 53.562 | 2.50E-13 |
| LC | 1 | 3 | 81.680 | 1.60E-19 |
| LC | 1 | 4 | 29.882 | 4.60E-08 |
| LC | 2 | 3 | 0.451 | 0.5 |
| LC | 2 | 4 | 77.738 | 1.20E-18 |
| LC | 3 | 4 | 115.643 | 5.70E-27 |
| Rigour3 | | | | |
| LC | 1 | 2 | 53.699 | 2.30E-13 |
| LC | 1 | 3 | 100.280 | 1.30E-23 |
| LC | 1 | 4 | 46.366 | 9.80E-12 |
| LC | 2 | 3 | 0.159 | 0.69 |

| Models for indicators | | | Wald | p |
|---|---|---|---|---|
| LC | 2 | 4 | 88.306 | 5.60E-21 |
| LC | 3 | 4 | 145.466 | 1.70E-33 |
| Rigour4 | | | | |
| LC | 1 | 2 | 29.856 | 4.70E-08 |
| LC | 1 | 3 | 45.023 | 1.90E-11 |
| LC | 1 | 4 | 22.655 | 1.90E-06 |
| LC | 2 | 3 | 0.216 | 0.64 |
| LC | 2 | 4 | 47.513 | 5.50E-12 |
| LC | 3 | 4 | 72.369 | 1.80E-17 |
| Rigour5 | | | | |
| LC | 1 | 2 | 35.716 | 2.30E-09 |
| LC | 1 | 3 | 43.549 | 4.10E-11 |
| LC | 1 | 4 | 17.542 | 2.80E-05 |
| LC | 2 | 3 | 0.058 | 0.81 |
| LC | 2 | 4 | 57.671 | 3.10E-14 |
| LC | 3 | 4 | 67.213 | 2.40E-16 |
| Rel1 | | | | |
| LC | 1 | 2 | 11.433 | 0.00072 |
| LC | 1 | 3 | 57.910 | 2.70E-14 |
| LC | 1 | 4 | 22.570 | 2.00E-06 |
| LC | 2 | 3 | 25.272 | 5.00E-07 |
| LC | 2 | 4 | 37.306 | 1.00E-09 |
| LC | 3 | 4 | 83.256 | 7.20E-20 |
| Rel2 | | | | |
| LC | 1 | 2 | 2.850 | 0.091 |
| LC | 1 | 3 | 27.982 | 1.20E-07 |
| LC | 1 | 4 | 19.730 | 8.90E-06 |
| LC | 2 | 3 | 31.765 | 1.70E-08 |
| LC | 2 | 4 | 8.027 | 0.0046 |
| LC | 3 | 4 | 55.061 | 1.20E-13 |
| KultErb | | | | |
| LC | 1 | 2 | 3.512 | 0.061 |
| LC | 1 | 3 | 22.144 | 2.50E-06 |
| LC | 1 | 4 | 12.749 | 0.00036 |
| LC | 2 | 3 | 31.989 | 1.60E-08 |
| LC | 2 | 4 | 3.230 | 0.072 |
| LC | 3 | 4 | 43.804 | 3.60E-11 |
| Komplex | | | | |
| LC | 1 | 2 | 0.232 | 0.63 |
| LC | 1 | 3 | 37.331 | 1.00E-09 |
| LC | 1 | 4 | 5.022 | 0.025 |
| LC | 2 | 3 | 31.529 | 2.00E-08 |
| LC | 2 | 4 | 2.629 | 0.1 |
| LC | 3 | 4 | 44.587 | 2.40E-11 |
| Vielfalt | | | | |
| LC | 1 | 2 | 0.796 | 0.37 |
| LC | 1 | 3 | 27.511 | 1.60E-07 |
| LC | 1 | 4 | 5.657 | 0.017 |
| LC | 2 | 3 | 27.068 | 2.00E-07 |
| LC | 2 | 4 | 2.349 | 0.13 |
| LC | 3 | 4 | 36.524 | 1.50E-09 |
| P_CV | | | | |
| LC | 1 | 2 | 8.285 | 0.004 |
| LC | 1 | 3 | 33.954 | 5.60E-09 |
| LC | 1 | 4 | 50.228 | 1.40E-12 |
| LC | 2 | 3 | 2.677 | 0.1 |

| Models for indicators | | | Wald | p |
|---|---|---|---|---|
| LC | 2 | 4 | 45.973 | 1.20E-11 |
| LC | 3 | 4 | 87.625 | 7.90E-21 |
| P_Dipl | | | | |
| LC | 1 | 2 | 6.147 | 0.013 |
| LC | 1 | 3 | 33.340 | 7.70E-09 |
| LC | 1 | 4 | 52.574 | 4.10E-13 |
| LC | 2 | 3 | 5.342 | 0.021 |
| LC | 2 | 4 | 49.398 | 2.10E-12 |
| LC | 3 | 4 | 88.464 | 5.20E-21 |
| P_PubList | | | | |
| LC | 1 | 2 | 6.940 | 0.0085 |
| LC | 1 | 3 | 32.230 | 1.40E-08 |
| LC | 1 | 4 | 35.958 | 2.00E-09 |
| LC | 2 | 3 | 4.773 | 0.029 |
| LC | 2 | 4 | 42.212 | 8.20E-11 |
| LC | 3 | 4 | 75.259 | 4.10E-18 |
| P_Letter | | | | |
| LC | 1 | 2 | 0.339 | 0.56 |
| LC | 1 | 3 | 17.294 | 3.20E-05 |
| LC | 1 | 4 | 50.894 | 9.70E-13 |
| LC | 2 | 3 | 8.287 | 0.004 |
| LC | 2 | 4 | 43.084 | 5.20E-11 |
| LC | 3 | 4 | 75.671 | 3.40E-18 |

Table 11. Wald statistics of the paired comparisons of the subclasses LC 1.1 and 1.2.

| Models for indicators | | | Wald | p |
|---|---|---|---|---|
| Eigenst | | | | |
| LC | 1.1 | 1.2 | 1.285 | 0.26 |
| Orig1 | | | | |
| LC | 1.1 | 1.2 | 0.000 | 0.98 |
| Orig2 | | | | |
| LC | 1.1 | 1.2 | 18.937 | 1.40E-05 |
| Orig3 | | | | |
| LC | 1.1 | 1.2 | 4.032 | 0.045 |
| Orig4 | | | | |
| LC | 1.1 | 1.2 | 18.361 | 1.80E-05 |
| Orig5 | | | | |
| LC | 1.1 | 1.2 | 15.286 | 9.20E-05 |
| Orig6 | | | | |
| LC | 1.1 | 1.2 | 5.155 | 0.023 |
| Real1 | | | | |
| LC | 1.1 | 1.2 | 0.026 | 0.87 |
| Real2 | | | | |
| LC | 1.1 | 1.2 | 0.000 | 0.99 |
| Rigour1 | | | | |
| LC | 1.1 | 1.2 | 0.459 | 0.5 |
| Rigour2 | | | | |
| LC | 1.1 | 1.2 | 1.145 | 0.28 |

| Models for indicators | | | Wald | p |
|---|---|---|---|---|
| Rigour3 | | | | |
| LC | 1.1 | 1.2 | 1.973 | 0.16 |
| Rigour4 | | | | |
| LC | 1.1 | 1.2 | 0.047 | 0.83 |
| Rigour5 | | | | |
| LC | 1.1 | 1.2 | 0.316 | 0.57 |
| Rel1 | | | | |
| LC | 1.1 | 1.2 | 0.499 | 0.48 |
| Rel2 | | | | |
| LC | 1.1 | 1.2 | 1.175 | 0.28 |
| KultErb | | | | |
| LC | 1.1 | 1.2 | 1.971 | 0.16 |
| Komplex | | | | |
| LC | 1.1 | 1.2 | 0.412 | 0.52 |
| Vielfalt | | | | |
| LC | 1.1 | 1.2 | 1.405 | 0.24 |
| P_CV | | | | |
| LC | 1.1 | 1.2 | 0.025 | 0.87 |
| P_Dipl | | | | |
| LC | 1.1 | 1.2 | 0.150 | 0.7 |
| P_PubList | | | | |
| LC | 1.1 | 1.2 | 0.155 | 0.69 |
| P_Letter | | | | |
| LC | 1.1 | 1.2 | 0.149 | 0.7 |

**Covariate analysis**

Table 12. Wald statistics of the covariates by main classes and subclasses of the final latent class tree (n = 760).

| Covariates | Main classes Wald | p | Subclasses Wald | p |
|---|---|---|---|---|
| Main effects | | | | |
| Gender | 10.202 | 0.017 | 3.251 | 0.071 |
| Application of doc or postdoc | 2.570 | 0.46 | 0.037 | 0.85 |
| Field | 4.236 | 0.64 | 6.589 | 0.037 |
| Age | 27.562 | 0.0065 | 3.488 | 0.48 |
| Tenure | 4874.095 | < 0.001 | 0.173 | 0.68 |
| Orientation of research | 6.862 | 0.33 | 0.279 | 0.87 |
| Experience as grant reviewer | 1.972 | 0.58 | 2.176 | 0.14 |
| Interaction effects | | | | |
| Age x Tenure | 5268.854 | < 0.001 | 2.725 | 0.6 |
| Age x Experience as grant reviewer | 10.380 | 0.58 | 1.945 | 0.75 |

Table 13. Wald statistics of the paired comparisons of the main classes LC 1, 2, 3, and 4 for the covariates gender, age, tenure and the interaction of age and tenure (n = 760).

| Paired comparison | Gender Wald | p | Age Wald | p | Tenure Wald | p | Age x Tenure Wald | p |
|---|---|---|---|---|---|---|---|---|
| LC 1 and 2 | 0.114 | 0.74 | 4.378 | 0.36 | 0.313 | 0.58 | 2.994 | 0.56 |
| LC 1 and 3 | 3.524 | 0.061 | 4.285 | 0.37 | 0.128 | 0.72 | 2.411 | 0.66 |
| LC 1 and 4 | 3.956 | 0.047 | 12.437 | 0.014 | 4873.741 | <0.001 | 4902.716 | <0.001 |
| LC 2 and 3 | 1.514 | 0.22 | 3.267 | 0.51 | 0.007 | 0.94 | 4.942 | 0.29 |
| LC 2 and 4 | 4.140 | 0.042 | 17.060 | 0.0019 | 2186.185 | <0.001 | 2146.367 | <0.001 |
| LC 3 and 4 | 9.807 | 0.0017 | 17.376 | 0.0016 | 1672.608 | <0.001 | 1677.032 | <0.001 |

Table 14. Relative frequency of the covariates in the sample (n = 760) and class specific probabilities of the covariates.

| Covariates | Sample | Main classes LC 1 | LC 2 | LC 3 | LC 4 | Subclasses LC 1.1 | LC 1.2 |
|---|---|---|---|---|---|---|---|
| **Gender** | | | | | | | |
| Men | 0.59 | 0.59 | 0.62 | 0.51 | 0.71 | 0.55 | 0.67 |
| Women | 0.41 | 0.41 | 0.38 | 0.49 | 0.29 | 0.45 | 0.33 |
| **Criteria for applications of** | | | | | | | |
| Doctoral students | 0.53 | 0.52 | 0.53 | 0.50 | 0.61 | 0.51 | 0.56 |
| Postdoctoral researchers | 0.47 | 0.48 | 0.47 | 0.50 | 0.39 | 0.49 | 0.44 |
| **Field** | | | | | | | |
| Languages and literatures | 0.40 | 0.43 | 0.42 | 0.34 | 0.39 | 0.50 | 0.31 |
| History and cultural sciences | 0.41 | 0.39 | 0.41 | 0.48 | 0.41 | 0.35 | 0.45 |
| Law | 0.18 | 0.18 | 0.17 | 0.19 | 0.21 | 0.15 | 0.24 |
| **Age** | | | | | | | |
| 28-37 | 0.19 | 0.22 | 0.15 | 0.13 | 0.22 | 0.21 | 0.26 |
| 38-43 | 0.21 | 0.22 | 0.09 | 0.21 | 0.40 | 0.22 | 0.22 |
| 44-50 | 0.21 | 0.19 | 0.29 | 0.23 | 0.11 | 0.20 | 0.18 |
| 51-57 | 0.19 | 0.17 | 0.23 | 0.27 | 0.18 | 0.15 | 0.18 |
| 58-82 | 0.20 | 0.20 | 0.25 | 0.17 | 0.09 | 0.22 | 0.16 |
| **Tenure** | | | | | | | |
| Yes | 0.38 | 0.34 | 0.52 | 0.39 | 0.26 | 0.35 | 0.34 |
| No | 0.62 | 0.66 | 0.48 | 0.61 | 0.74 | 0.65 | 0.66 |
| **Orientation of research** | | | | | | | |
| Mainly national | 0.14 | 0.18 | 0.12 | 0.10 | 0.12 | 0.15 | 0.20 |
| Mainly international | 0.40 | 0.39 | 0.45 | 0.41 | 0.38 | 0.38 | 0.39 |
| Equally national and international | 0.45 | 0.44 | 0.44 | 0.49 | 0.50 | 0.47 | 0.41 |
| **Experience as grant reviewer** | | | | | | | |
| Yes | 0.55 | 0.49 | 0.70 | 0.60 | 0.47 | 0.47 | 0.55 |
| No | 0.45 | 0.51 | 0.30 | 0.40 | 0.53 | 0.53 | 0.45 |